\documentclass[12pt,a4paper]{article}
\usepackage{graphicx}
\usepackage{subeqnarray}

\begin{document}
%
%
%
\newenvironment{lefteqnarray}{\arraycolsep=0pt\begin{eqnarray}}
{\end{eqnarray}\protect\aftergroup\ignorespaces}
\newenvironment{lefteqnarray*}{\arraycolsep=0pt\begin{eqnarray*}}
{\end{eqnarray*}\protect\aftergroup\ignorespaces}
\newenvironment{leftsubeqnarray}{\arraycolsep=0pt\begin{subeqnarray}}
{\end{subeqnarray}\protect\aftergroup\ignorespaces}
\newcommand{\appleq}{\stackrel{<}{\sim}}
\newcommand{\appgeq}{\stackrel{>}{\sim}}
\newcommand{\arcsinh}{\mathop{\rm arcsinh}\nolimits}
\newcommand{\arctg}{\mathop{\rm arctg}\nolimits}
\newcommand{\diff}{{\rm\,d}}
\newcommand{\determinant}[1]{\left|\matrix{#1}\right|}
\newcommand{\displayfrac}[2]{\frac{\displaystyle #1}{\displaystyle #2}}
\newcommand{\Erfc}{\mathop{\rm Erfc}\nolimits}
\newcommand{\Int}{\mathop{\rm Int}\nolimits}
\newcommand{\Nint}{\mathop{\rm Nint}\nolimits}
\newcommand{\odiv}{\mathop{\rm div}\nolimits}
\newcommand{\pprime}{{\prime\prime}}

\title{Pontential energies and potential-energy tensors for subsystems: 
        general properties}
\author{{R. Caimmi\footnote{
{\it Physics and Astronomy Department, Padua University,
Vicolo Osservatorio 3/2, I-35122 Padova, Italy.  Affiliated up to September
30th 2014.   Current status: Studioso Senior.  Current position: in retirement
due to age limits.}\hspace{50mm}
email: roberto.caimmi@unipd.it~~~
fax: 39-049-8278212
}
\phantom{agga}}}



\maketitle
\begin{quotation}
\section*{}
\begin{Large}
\begin{center}

Abstract

\end{center}
\end{Large}
\begin{small}

With regard to generic two-component systems, the theory of first variations
of global quantities is reviewed and explicit expressions are inferred for
subsystem potential energies and potential-energy tensors.   Performing a
conceptual experiment, a physical interpretation of subsystem potential
energies and potential-energy tensors is discussed.   Subsystem tidal radii 
are defined by requiring an unbound component in absence of the other one.
To this respect, a few guidance examples are presented as:
(i) an embedding and an embedded homogeneous sphere; (ii) an embedding and an
embedded truncated, singular isothermal sphere where related centres are
sufficiently distant; (iii) a homogeneous sphere and a Roche system i.e. a
mass point surrounded by a vanishing atmosphere.   The results are discussed
and compared with the findings of earlier investigations.

\noindent
{\it keywords - gravitation: Newtonian theory; virial theorem; cosmology:
dark matter; galaxies: tidal radius; galaxies; globular clusters: tidal
radius.}

\end{small}
\end{quotation}

%

\section{Introduction} \label{intro}

Potential energies and potential-energy tensors are key ingredients for the
application of the virial method (Chandrasekhar 1969, hereafter quoted as C69,
Chap.\,2), that is
essentially the method of the moments applied to the solution of
hydrodynamical problems in which the gravitational field of the prevailing
distribution of matter is taken into account.   In particular, the first
variations of potential energies and potential-energy tensors caused by a
perturbation (C69, Chap.\,2, \S15) are needed for a number of
applications, such as practical use of virial equations in linearized form
for the treatment of the stability of a configuration (C69, Chap.\,3, \S23),
the effect of viscous dissipation on the stability (C69, Chap.\,5, \S37;
Chap.\,8, \S59), the determination of bifurcation points (C69, Chap.\,6, \S45)
and loci of neutral points belonging to third harmonics (C69, Chap.\,7, \S50).

Large-scale celestial bodies e.g., galaxies and galaxy clusters,
appear to be made of at least two  subsystems which link only via
gravitational interaction, where each component is distorted by the tidal
force induced by the remaining one(s).   On the other hand, large-scale
celestial bodies can
no longer be conceived as isolated and are often sufficiently close to exhibit
tidal effects even in absence of accretion or merging.

The formulation of the virial theorem (implying the application of the virial
method) to each subsystem separately yields a
larger amount of information with respect to the system as a whole (Limber
1959; Brosche et al. 1983; Caimmi et al. 1984).   To this respect, different
kinds of potential energies and potential-energy tensors can be defined (e.g.,
Caimmi and Secco 1992), namely (i) self, related to the integration of the
gravitational potential from the subsystem under consideration on the mass
distribution of the subsystem under consideration; (ii) interaction, related
to the integration of the
gravitational potential from another subsystem on the mass distribution of the
subsystem under consideration; (iii) tidal, related to the integration of the
virial due to the gravitational force
from another subsystem on the mass distribution of the subsystem under
consideration; (iv) residual, which is merely the difference tidal minus
interaction.

Potential energies and potential-energy tensors of subsystems can be
used, among others, for a definition of tidal radius (Secco 2000; Caimmi and
Secco 2003), an interpretation of the fundamental plane of elliptical galaxies
(Secco 2001), a formulation of stellar system thermodynamics (Secco 2005), and
related first variations can be used for an application of d'Alembert's
principle involving the determination of virtual displacements (Secco 2001,
2005).

The current paper is restricted to two-component systems, without loss of
generality in that multi-component systems can be conceived as the collection
of all the pairs made of a selected subsystem and another one.   With regard
to a generic two-component system, investigation is devoted to the following
points: explicit expression of first variations of potential energies and
potential-energy tensors with the addition of physical interpretation,
considered in Section \ref{fiva} and \ref{phin},
respectively; a global criterion for the definition of tidal radius,
considered in Section \ref{trats}, where a few guidance examples are
presented.   The discussion and the conclusion make the subject of
Section \ref{disc} and \ref{conc}, respectively.   Further details on a number
of arguments mentioned in the text are shown in the Appendix.

\section{First variations}
\label{fiva}
\subsection{General remarks}
\label{gere}

Let an unperturbed (collisional or
collisionless) self-gravitating fluid be taken into consideration,
filling the volume, $S_o$,
at the time, $t$, and let $\Phi_o$ be a global parameter, 
depending on a local parameter, ${\cal Q}_o$, as:
\begin{equation}
\Phi_o=\int_{S_o}{\cal Q}_o(x_{o1}, x_{o2}, x_{o3}, t)\diff^3S_o~~;
\label{eq:phio}
\end{equation}
where $\diff^3S_o=\diff x_{o1}\diff x_{o2}\diff x_{o3}$ is the
volume of an infinitesimal (unperturbed) fluid element.

If the fluid has occurred to be slightly perturbed at some
initial time, $t_i<t$, a different evolution takes place
from that time on, and the global parameter under
consideration reads:
\begin{equation}
\Phi=\int_{S}{\cal Q}(x_{1}, x_{2}, x_{3}, t)\diff^3S~~;
\label{eq:phi}
\end{equation}
where $\diff^3S=\diff x_{1}\diff x_{2}\diff x_{3}$ is the
volume of an infinitesimal (perturbed) fluid element.

According to the parent paper (C69, Chap.\,2, \S\,15), the first
variation of the global parameter, $\Phi$, caused by the
perturbation, is defined as:
\begin{equation}
\delta\Phi=
\int_{S}{\cal Q}(\vec{r}, t)\diff^3S
-\int_{S_o}{\cal Q}_o(\vec{r}_o, t)\diff^3S_o~~;
\label{eq:dphi}
\end{equation}
where  $\vec{r}\equiv(x_{1}, x_{2}, x_{3})$, $\vec{r}_o\equiv(x_{o1},
x_{o2}, x_{o3})$, and the coordinates of perturbed fluid
elements are related
to their unperturbed counterparts by the transformation:
\begin{leftsubeqnarray}
\slabel{eq:xka}
&& x_k=x_{ok}+\xi_k(\vec{r}_o,t)~~; \\
\slabel{eq:xkb}
&& \left\vert\frac{x_k-x_{ok}}{x_{ok}}\right\vert\ll1~~;
\qquad k=1,2,3~~;
\label{seq:xk}
\end{leftsubeqnarray}
or, in other words, the perturbed fluid is in linear regime.

A change of variables, defined by Eq.\,(\ref{eq:xka}),
implies the Jacobian:
\begin{lefteqnarray}
\label{eq:J}
&& J(x_{o1}, x_{o2}, x_{o3},t)  \nonumber \\
&& \nonumber \\
&& = \determinant{\displayfrac{\partial x_1}{\partial x_{o1}} 
& \displayfrac{\partial x_2}{\partial x_{o1}} & 
\displayfrac{\partial x_3}{\partial x_{o1}} \cr
\displayfrac{\partial x_1}{\partial x_{o2}} & 
\displayfrac{\partial x_2}{\partial x_{o2}} & 
\displayfrac{\partial x_3}{\partial x_{o2}} \cr
\displayfrac{\partial x_1}{\partial x_{o3}} & 
\displayfrac{\partial x_2}{\partial x_{o3}} & 
\displayfrac{\partial x_3}{\partial x_{o3}}
\cr} = \determinant{
1+\displayfrac{\partial \xi_1}{\partial x_{o1}} &
  \displayfrac{\partial \xi_2}{\partial x_{o1}} &
  \displayfrac{\partial \xi_3}{\partial x_{o1}} \cr
  \displayfrac{\partial \xi_1}{\partial x_{o2}} &
1+\displayfrac{\partial \xi_2}{\partial x_{o2}} &
  \displayfrac{\partial \xi_3}{\partial x_{o2}} \cr
  \displayfrac{\partial \xi_1}{\partial x_{o3}} &
  \displayfrac{\partial \xi_2}{\partial x_{o3}} &
1+\displayfrac{\partial \xi_3}{\partial x_{o3}}
\cr}~~;
\end{lefteqnarray}
which may safely be approximated as:
\begin{lefteqnarray}
\label{eq:Ja}
&& J(x_{o1},x_{o2},x_{o3},t)=1+\odiv\vec{\xi}~~;
\end{lefteqnarray}
where $\vec{\xi}\equiv[\xi_1(\vec{r}_o,t),\xi_2(\vec{r}_o,t),\xi_3
(\vec{r}_o,t)]$, to the first order in the displacement.

Accordingly, the first variation, $\delta\Phi$,
expressed by Eq.\,(\ref{eq:dphi}), reads:
\begin{lefteqnarray*}
\label{eq:dpq}
&& \delta\Phi=\int_{S_o}{\cal Q}(\vec{r}_o+
\vec{\xi},t)(1+\odiv\vec{\xi})
\diff^3S_o
-\int_{S_o}{\cal Q}_o
(\vec{r}_o,t) \diff^3S_o \nonumber \\
&& \phantom{\delta\Phi}=\int_{S_o}{\cal Q}
(\vec{r}_o+\vec{\xi},t)\diff^3S_o
+\int_{S_o}{\cal Q}
(\vec{r}_o+\vec{\xi},t)\odiv\vec{\xi}\diff^3S_o
-\int_{S_o}{\cal Q}_o(\vec{r}_o,t) \diff^3S_o~~;
\end{lefteqnarray*}
which is equivalent to:
\begin{lefteqnarray}
\label{eq:dpdq}
&& \delta\Phi=\int_{S_o}(\Delta{\cal Q}+{\cal Q}\odiv
\vec{\xi})\diff^3S_o~~;
\end{lefteqnarray}
where $\Delta{\cal Q}={\cal Q}(\vec{r}_o+\vec{\xi},t)-{\cal Q}_o(\vec{r}_o,t)$
is the Lagrangian change in
the local parameter, ${\cal Q}$, consequent to the
displacement, $\vec{\xi}$ (C69,
Chap.\,2, \S\,13).

The particularization of Eq.\,(\ref{eq:dpdq}) to the
special case where the local parameter coincides
with the density i.e. ${\cal Q}=\rho$, yields:
\begin{lefteqnarray}
\label{eq:dpdr}
&& \delta\Phi=\delta\int_{S_o}\rho\diff^3S_o=\int_{S_o}
(\Delta\rho+\rho\odiv\vec{\xi})\diff^3S_o=0~~;
\end{lefteqnarray}
owing to mass conservation during the first
variation (C69, Chap.\,2, \S\,15).

The particularization of Eq.\,(\ref{eq:dpdq}) to the
special case where the local parameter is expressible
as a product where a factor is the density and the
other one is an additional local parameter i.e. ${\cal
Q}^\prime=\rho{\cal Q}$, yields:
\begin{lefteqnarray*}
&& \delta\Phi=\delta\int_{S_o}{\cal Q}^\prime\diff^3S_o
=\delta
\int_{S_o}\rho{\cal Q}\diff^3S_o=\int_{S_o}[\Delta(\rho
{\cal Q})+\rho{\cal Q}\odiv\vec{\xi}]\diff^3S_o~~;
\end{lefteqnarray*}
which reduces to:
\begin{lefteqnarray}
\label{eq:dpdqr}
&& \delta\Phi=\delta\int_{S_o}\rho{\cal Q}\diff^3S_o=
\int_{S_o}\rho\Delta{\cal Q}\diff^3S_o~~;
\end{lefteqnarray}
owing to Eq.\,(\ref{eq:dpdr}).

The further restriction that the local parameter,
${\cal Q}$, is not intrinsic to a generic fluid
element, such as pressure or density, but something
which it assumes simply by virtue of its position,
such as gravitational potential, allows
the validity of the relation (C69,
Chap.\,2, \S\,13):
\begin{lefteqnarray}
\label{eq:DQ}
&& \Delta{\cal Q}=\sum_{k=1}^3\xi_k\frac
{\partial{\cal Q}}{\partial x_{ok}}~~;
\end{lefteqnarray}
and Eq.\,(\ref{eq:dpdqr}) takes the form:
\begin{lefteqnarray}
\label{eq:dFS}
&& \delta\Phi=\delta\int_{S_o}\rho{\cal Q}\diff^3S_o=
\int_{S_o}\rho\sum_{k=1}^3\xi_k\frac{\partial{\cal Q}}
{\partial x_{ok}}\diff^3S_o~~;
\end{lefteqnarray}
if the global parameter, $\Phi$, is a vector, or a
tensor, then a similar relation holds for the first
variation of each component.

The generalization of Eq.\,(\ref{eq:dFS}) to the case
where the local parameter is expressible as a product,
two factors being densities calculated at different
points, and a third factor being an additional local
parameter (not intrinsic to a generic fluid element)
which depends on both positions i.e.
${\cal Q}^\prime(\vec{r},\vec{r^\prime},t)=\rho(\vec
{r},t)\rho(\vec{r^\prime},t){\cal Q}(\vec{r},\vec{r^
\prime},t)$, reads (C69, Chap.\,2, \S\,15):
\begin{lefteqnarray}
\label{eq:dF12}
&& \delta\Phi=\delta\int_{S_{o}}\int_{S_{o}}\rho
(\vec{r}_o,t)\rho(\vec{r^\prime}_o,t){\cal Q}(\vec
{r}_o,\vec{r^\prime}_o,t)\diff^3S_o\diff^3S_o^\prime
\nonumber \\
&&
=\int_{S_o}\int_{S_o}\rho
(\vec{r}_o,t)\rho(\vec{r^\prime}_o,t)\sum_{k=1}
^3\left[\xi_k(\vec{r}_o,t)\frac{\partial{\cal Q}}
{\partial x_{ok}}+\xi_k(\vec{r^\prime}_o,t)\frac
{\partial{\cal Q}}{\partial x_{ok}^\prime}\right]
\diff^3S_o\diff^3S_o^\prime~~;\qquad
\end{lefteqnarray}
where $\diff^3S_o$, $\diff^3S_o^\prime$, are
infinitesimal volume elements on the top of
the radius vector, $\vec{r}_o$, $\vec{r^\prime}_o$,
respectively.
If the global parameter, $\Phi$, is a vector, or a
tensor, then a similar relation holds for the first
variation of each component.

The generalization of Eq.\,(\ref{eq:dF12}) to the case
where the local parameter is expressible as a product,
two factors being densities calculated at different
points of different subsystems, denoted as $u$, $v$,
respectively, and a third factor being an additional
local parameter (not intrinsic to a generic fluid
element) which depends on both positions i.e.
${\cal Q}^\prime(\vec{r}_u,\vec{r}_v,t)=\rho_u(\vec
{r}_u,t)\rho_v(\vec{r}_v,t){\cal Q}(\vec{r}_u,\vec{r}_
v,t)$, reads:
\begin{lefteqnarray}
\label{eq:dF2}
&& \delta\Phi=\delta\int_{S_{ou}}\int_{S_{ov}}
\rho_u(\vec{r}_{ou},t)\rho_v(\vec{r}_{ov},t)
{\cal Q}(\vec{r}_{ou},\vec{r}_{ov},t)
\diff^3S_{ou}\diff^3S_{ov} \nonumber \\
&& \phantom{\delta\Phi}=\int_{S_{ou}}\int_{S_{ov}}
\rho_u(\vec{r}_{ou},t)\rho_v(\vec{r}_{ov},t)
\nonumber \\
&& \phantom{\delta\Phi=}\times
\sum_{k=1}^3\left[\xi_k^{(u)}(\vec{r}_{ou},t)
\frac{\partial{\cal Q}}{\partial x_{ok}^{(u)}}
+\xi_k^{(v)}(\vec{r}_{ov},t)\frac{\partial{\cal
Q}}{\partial x_{ok}^{(v)}}\right]\diff^3S_{ou}
\diff^3S_{ov}~~;
\end{lefteqnarray}
where $\diff^3S_{ou}$, $\diff^3S_{ov}$, are
infinitesimal volume elements on the top of
the radius vector, $\vec{r}_{ou}$, $\vec{r}
_{ov}$, respectively, and
$\vec{r}_{ow}\equiv(x_{o1}^{(w)},x_{o2}^{(w)},x_{o3}^{(w)})$, $w=u,v$.
If the global parameter, $\Phi$, is a vector, or a
tensor, then a similar relation holds for the first
variation of each component.
%

\subsection{Potential-energy
tensors for subsystems}\label{pote}

Let an unperturbed (collisional or collisionless), two-component,
self-gravitating fluid be taken into consideration, where the subsystems,
denoted as $i$ and $j$, respectively, interact
only gravitationally.   In finding the first
variations of global parameters, let attention
be restricted to the potential-energy tensors
(C69, Chap.\,2, \S\,10; Brosche et al.
1983; Caimmi et al. 1984; Caimmi and Secco 1992):
\begin{lefteqnarray}
\label{eq:Oupq}
&& (\Omega_u)_{pq}=-\frac12\int_{S_u}\rho_u(\vec
{r})({\cal V}_u)_{pq}(\vec{r})\diff^3S_u~~; \\
\label{eq:Wuvpq}                                     
&& (W_{uv})_{pq}=-\frac12\int_{S_u}\rho_u(\vec
{r})({\cal V}_v)_{pq}(\vec{r})\diff^3S_u~~; \\
\label{eq:Vuvpq}                                     
&& (V_{uv})_{pq}=\int_{S_u}\rho_u(\vec{r})x_p
\frac{\partial{\cal V}_v}{\partial x_q}\diff^3S_u~~;
\end{lefteqnarray}
where $u=i,j$; $v=j,i$; $\diff^3S_u=\diff x_1\diff x_2\diff x_3$;
$({\cal V}_u)_{pq}$
and ${\cal V}_u$ are the tensor potential
and the potential, respectively (C69,
Chap.\,2, \S\,10):
\begin{lefteqnarray}
\label{eq:Pupq}
&& ({\cal V}_u)_{pq}(\vec{r})=G\int_{S_u}\rho_u(\vec
{r^\prime})\displayfrac{(x_p-x_p^\prime)
(x_q-x_q^\prime)}{\left[\sum_{k=1}^3(x_k-x_k^\prime)^2
\right]^{3/2}}\diff^3S_u^\prime~~; \\
\label{eq:Pu}
&& {\cal V}_u(\vec{r})=G\int_{S_u}\rho_u(\vec{r^\prime})
\left[\sum_{k=1}^3(x_k-x_k^\prime)^2\right]^{-1/2}
\diff^3S_u^\prime~~; \\
\label{eq:dPdx}                                     
&& \frac{\partial{\cal V}_u}{\partial x_p}=G\int_{S_u}
\rho_u(\vec{r^\prime})\frac{\partial}
{\partial x_p}\left[\sum_{k=1}^3(x_k-x_k^\prime)^2
\right]^{-1/2}\diff^3S_u^\prime \nonumber \\
&& \phantom{\frac{\partial{\cal V}_u}{\partial x_p}}
=-G\int_{S_u}\rho_u(\vec{r^\prime})\displayfrac
{x_p-x_p^\prime}{\left[\sum_{k=1}^3(x_k-x_k^\prime)^2
\right]^{3/2}}\diff^3S_u^\prime~~;
\end{lefteqnarray}
where $G$ is the constant of gravitation and
$\diff^3S_u^\prime=\diff x_1^\prime\diff x_2^\prime\diff x_3^\prime$.

The gravitational potential, ${\cal V}_u$, the
potential self energy, $\Omega_u$, the potential
interaction energy, $W_{uv}$, and the potential
tidal energy, $V_{uv}$, make the trace of their
tensor counterparts (C69, Chap.\,2,
\S\,10; Brosche et al. 1983; Caimmi et al. 1984;
Caimmi and Secco 1992):
\begin{equation}
\label{eq:tens}
\sum_{r=1}^3\Psi_{rr}=\Psi~~;\qquad\Psi=
{\cal V}_u,\Omega_u, W_{uv}, V_{uv}~~;
\end{equation}
for a formal demonstration, an interested reader
is addressed to the above quoted parent papers.

Let the local parameter be taken equal to the
integrand, without density factors, in the explicit expression
of the potential self-energy tensor, defined
by Eq.\,(\ref{eq:Oupq}) via (\ref{eq:Pupq}).   Using Eq.\,(\ref{eq:dF12}),
the first variation of the potential self-energy
tensor after some algebra reads (C69, Chap.\,2, \S\,15):
\begin{equation}
\delta(\Omega_u)_{pq}=-\int_{S_{ou}}\rho_u
\sum_{k=1}^3\left[\xi_k^{(u)}\frac{\partial({\cal V}_u)_
{pq}}{\partial x_k^{(u)}}\right]\diff^3S_{ou}~~;
\label{eq:dOpq}
\end{equation}
and the trace of the above tensor, owing to
Eq.\,(\ref{eq:tens}), reads:
\begin{equation}
\label{eq:dO}
\delta\Omega_u=-\int_{S_{ou}}\rho_u
\sum_{k=1}^3\left[\xi_k^{(u)}\frac{\partial{\cal V}_u}
{\partial x_k^{(u)}}\right]\diff^3S_{ou}~~;
\end{equation}
that is the first variation of the potential
self energy.

Let the local parameter be taken equal to the
integrand, without density factors, in the explicit expression
of the potential interaction-energy tensor, defined
by Eq.\,(\ref{eq:Wuvpq}) via (\ref{eq:Pupq}).   Using Eq.\,(\ref{eq:dF2}),
the first variation of the potential interaction-energy
tensor after some algebra reads:
\begin{lefteqnarray}
\label{eq:dWpq}
&& \delta(W_{uv})_{pq}=-\frac12\int_{S_{ou}}\rho_u
\sum_{k=1}^3\left[\xi_k^{(u)}\frac{\partial({\cal V}_v)_
{pq}}{\partial x_k^{(u)}}\right]\diff^3S_{ou} \nonumber \\
&& \phantom{\delta(W_{uv})_{pq}=}-\frac12\int_{S_
{ov}}\rho_v\sum_{k=1}^3\left[\xi_k^{(v)}\frac{\partial({\cal
V}_u)_{pq}}{\partial x_k^{(v)}}\right]\diff^3S_{ov}~~;
\end{lefteqnarray}
and the trace of the above tensor, owing to
Eq.\,(\ref{eq:tens}), reads:
\begin{lefteqnarray}
\label{eq:dW}
&& \delta W_{uv}=-\frac12\int_{S_{ou}}\rho_u
\sum_{k=1}^3\left[\xi_k^{(u)}\frac{\partial{\cal V}_v}
{\partial x_k^{(u)}}\right]\diff^3S_{ou} \nonumber \\
&& \phantom{\delta W_{uv}=}-\frac12\int_{S_{ov}}
\rho_v\sum_{k=1}^3\left[\xi_k^{(v)}\frac{\partial{\cal
V}_u}{\partial x_k^{(v)}}\right]\diff^3S_{ov}~~;
\end{lefteqnarray}
that is the first variation of the potential
interaction energy.

The sum of the first and the last term on
the right-hand side of Eqs.\,(\ref{eq:dWpq})
and (\ref{eq:dW}) is symmetric with respect
to the exchange of the indexes, $u$ and $v$,
which makes the following relations hold:
\begin{lefteqnarray}
\label{eq:dWijpq}
&& \delta (W_{ij})_{pq}+\delta (W_{ji})_{pq}
=-\int_{S_{oi}}
\rho_i\sum_{k=1}^3\left[\xi_k^{(i)}\frac{\partial({\cal V}_
j)_{pq}}{\partial x_k^{(i)}}\right]\diff^3S_{oi} \nonumber \\
&& \phantom{\delta (W_{ij})_{pq}+\delta (W_{ij})
_{pq}=}-\int_{S_{oj}}
\rho_j\sum_{k=1}^3\left[\xi_k^{(j)}\frac{\partial({\cal V}_i)
_{pq}}{\partial x_k^{(j)}}\right]\diff^3S_{oj}~~; \\
\label{eq:dWij}
&& \delta W_{ij}+\delta W_{ji}
=-\int_{S_{oi}}\rho_i
\sum_{k=1}^3\left[\xi_k^{(i)}\frac{\partial{\cal V}_j}
{\partial x_k^{(i)}}\right]\diff^3S_{oi} \nonumber \\
&& \phantom{\delta W_{ij}+\delta W_{ij}=}-\int_{S_{oj}}
\rho_j\sum_{k=1}^3\left[\xi_k^{(j)}\frac{\partial{\cal V}_i}
{\partial x_k^{(j)}}\right]\diff^3S_{oj}~~;
\end{lefteqnarray}
and, in addition:
\begin{lefteqnarray}
\label{eq:dWijpqs}
&& \delta (W_{ij})_{pq}=\delta (W_{ji})_{pq}~~; \\
\label{eq:dWijs}
&& \delta W_{ij}=\delta W_{ji}~~;
\end{lefteqnarray}
as expected from the symmetry of the potential
interaction-energy tensors with respect to the
exchange of the indexes, $i$ and $j$ (e.g.,
Caimmi and Secco 1992).

Let the local parameter be taken equal to the
integrand, without density factors, in the explicit expression
of the potential tidal-energy tensor, defined
by Eq.\,(\ref{eq:Vuvpq}) via (\ref{eq:dPdx}).   Using Eq.\,(\ref{eq:dF2}),
the first variation of the potential tidal-energy
tensor after some algebra reads:
\begin{lefteqnarray}
\label{eq:dVpq}
&& \delta(V_{uv})_{pq}=+\int_{S_{ou}}\rho_u
\sum_{k=1}^3\left\{\xi_k^{(u)}\frac{\partial}{\partial x_k^{(u)}}
\left[x_p^{(u)}\frac{\partial{\cal V}_v}{\partial x_q^{(u)}}
\right]\right\}\diff^3S_{ou} \nonumber \\
&& \phantom{\delta(V_{uv})_{pq}=}-\int_{S_
{ov}}\rho_v\sum_{k=1}^3\left[\xi_k^{(v)}\frac{\partial({\cal
V}_u)_{pq}}{\partial x_k^{(v)}}\right]\diff^3S_{ov}
\nonumber \\
&& \phantom{\delta(V_{uv})_{pq}=}-\int_{S_
{ov}}\rho_v\sum_{k=1}^3\left\{\xi_k^{(v)}\frac{\partial}
{\partial x_k^{(v)}}
\left[x_p^{(v)}\frac{\partial{\cal V}_u}
{\partial x_q^{(v)}}\right]\right\}\diff^3S_{ov}~~;
\end{lefteqnarray}
and the trace of the above tensor, owing to
Eq.\,(\ref{eq:tens}), reads:
\begin{lefteqnarray}
\label{eq:dV}
&& \delta V_{uv}=+\int_{S_{ou}}\rho_u
\sum_{k=1}^3\sum_{r=1}^3\left\{\xi_k^{(u)}\frac{\partial}
{\partial x_k^{(u)}}\left[x_r^{(u)}\frac{\partial{\cal V}_v}
{\partial x_r^{(u)}}\right]\right\}\diff^3S_{ou} \nonumber \\
&& \phantom{\delta V_{uv}=}-\int_{S_{ov}}
\rho_v\sum_{k=1}^3\left[\xi_k^{(v)}\frac{\partial{\cal V}_u}
{\partial x_k^{(v)}}\right]\diff^3S_{ov} \nonumber \\
&& \phantom{\delta V_{uv}=}-\int_{S_{ov}}
\rho_v\sum_{k=1}^3\sum_{r=1}^3\left\{\xi_k^{(v)}\frac
{\partial}{\partial x_k^{(v)}}\left[x_r^{(v)}\frac
{\partial{\cal V}_u}{\partial x_r^{(v)}}\right]\right\}
\diff^3S_{ov}~~;
\end{lefteqnarray}
that is the first variation of the potential
tidal energy.

The first term on the right-hand side of
Eqs.\,(\ref{eq:dVpq}) and (\ref{eq:dV})
is related to the effect of the variation
on $u$ subsystem, while the other two
terms are related to the effect of the
variation on $v$ subsystem.   In addition,
the sum of the first and the last term is
antisymmetric with respect to the exchange
of the indexes, $u$ and $v$, which makes
the following relations hold:
\begin{lefteqnarray}
\label{eq:dVijpq}
&& \delta (V_{ij})_{pq}+\delta (V_{ji})_{pq}
=-\int_{S_{oi}}\rho_i
\sum_{k=1}^3\left[\xi_k^{(i)}\frac{\partial({\cal V}_j)_
{pq}}{\partial x_k^{(i)}}\right]\diff^3S_{oi} \nonumber \\
&& \phantom{\delta (V_{ij})_{pq}+\delta (V_{ij})_
{pq}=}-\int_{S_{oj}}
\rho_j\sum_{k=1}^3\left[\xi_k^{(j)}\frac{\partial({\cal V}_
i)_{pq}}{\partial x_k^{(j)}}\right]\diff^3S_{oj}~~; \\
\label{eq:dVij}
&& \delta V_{ij}+\delta V_{ji}
=-\int_{S_{oi}}\rho_i
\sum_{k=1}^3\left[\xi_k^{(i)}\frac{\partial{\cal V}_j}
{\partial x_k^{(i)}}\right]\diff^3S_{oi} \nonumber \\
&& \phantom{\delta V_{ij}~+\delta (V_{ij})}-\int_{S_{oj}}
\rho_j\sum_{k=1}^3\left[\xi_k^{(j)}\frac{\partial{\cal V}_i}
{\partial x_k^{(j)}}\right]\diff^3S_{oj}~~;
\end{lefteqnarray}
and, in addition:
\begin{lefteqnarray}
\label{eq:dVijpqs}
&& \delta (V_{ij})_{pq}+\delta (V_{ji})_{pq}=
\delta (W_{ij})_{pq}+\delta (W_{ji})_{pq}~~; \\
\label{eq:dQijpqs}
&& \delta (Q_{ij})_{pq}+\delta (Q_{ji})_{pq}=0~~; \\
\label{eq:dVijs}
&& \delta V_{ij}+\delta V_{ji}=
\delta W_{ij}+\delta W_{ji}~~; \\
\label{eq:dQijs}
&& \delta Q_{ij}+\delta Q_{ji}=0~~;
\end{lefteqnarray}
as expected from the symmetry of the potential
interaction-energy tensors and the antisymmetry
of the potential residual-energy tensors:
\begin{eqnarray}
\label{eq:Quvpq}
&& (Q_{uv})_{pq}=(V_{uv})_{pq}-(W_{uv})_{pq}~~; \\
\label{eq:Q1}
&& Q_{uv}=V_{uv}-W_{uv}~~;
\end{eqnarray}
with respect to the exchange of the indexes,
$u$ and $v$, which translates into the following relations:
\begin{lefteqnarray}
\label{eq:Wst}
&& (W_{ij})_{pq}=(W_{ji})_{pq}~~; \\
\label{eq:Qat}
&& (Q_{ij})_{pq}=-(Q_{ji})_{pq}~~; \\
\label{eq:Ws}
&& W_{ij}=W_{ji}~~; \\
\label{eq:Qa}
&& Q_{ij}=-Q_{ji}~~;
\end{lefteqnarray}
for further details, an interested reader is addressed to the parent paper
(Caimmi and Secco 1992).

The combination of Eqs.\,(\ref{eq:dVijs}) and
(\ref{eq:dQijs}) yields:
\begin{lefteqnarray}
\label{eq:dVWQpqs}
&& \delta (V_{uv})_{pq}=
\delta (W_{uv})_{pq}+\delta (Q_{uv})_{pq}~~; \\
\label{eq:dVWQs}
&& \delta V_{uv}=\delta W_{uv}+\delta Q_{uv}~~;
\end{lefteqnarray}
via Eqs.\,(\ref{eq:Quvpq}) and (\ref{eq:Q1}).

A similar result holds for the potential
self-energy tensor of the whole system:
\begin{lefteqnarray}
\label{eq:OTpq}
&& \Omega_{pq}=(\Omega_i)_{pq}+(\Omega_j)_{pq}+
(W_{ij})_{pq}+(W_{ji})_{pq} \nonumber \\
&& \phantom{\Omega_{pq}}=(\Omega_i)_{pq}+(\Omega_
j)_{pq}+(V_{ij})_{pq}+(V_{ji})_{pq}~~;
\end{lefteqnarray}
where the related first variation, according
to Eqs.\,(\ref{eq:dF12}) and (\ref{eq:dOpq}), after some algebra reads:
\begin{lefteqnarray}
\label{eq:dOTpq}
&& \delta\Omega_{pq}=-\int_{S_{o}}\rho
\sum_{k=1}^3\xi_k\frac{\partial{\cal V}_
{pq}}{\partial x_k}\diff^3S_o  \nonumber \\
&& \phantom{\delta\Omega_{pq}}=-\int_{S_{o}}
(\rho_i+\rho_j)\sum_{k=1}^3\left[\xi_k^{(i)}\frac
{\partial({\cal V}_i)_{pq}}{\partial x_k^{(i)}}+\xi_
k^{(j)}\frac{\partial({\cal V}_j)_{pq}}{\partial x_k^
{(j)}}\right]\diff^3S_o~~;  \\
\end{lefteqnarray}
owing to the additivity of densities and
tensor potentials.   Splitting in four the
last integral, and using Eqs.\,(\ref{eq:dOpq}),
(\ref{eq:dWpq}), and
(\ref{eq:dVijs}), the final result is:
\begin{lefteqnarray}
\label{eq:dOWVpq}
&& \delta\Omega_{pq}=\delta(\Omega_i)_{pq}+
\delta(\Omega_j)_{pq}+\delta(W_{ij})_{pq}+
\delta(W_{ji})_{pq} \nonumber \\
&& \phantom{\delta\Omega_{pq}}=\delta(\Omega_
i)_{pq}+\delta(\Omega_j)_{pq}+\delta(V_{ij})
_{pq}+\delta(V_{ji})_{pq}~~;
\end{lefteqnarray}
and a summation over all the diagonal
components yields:
\begin{lefteqnarray}
\label{eq:dOWV}
&& \delta\Omega=\delta\Omega_i+
\delta\Omega_j+\delta W_{ij}+
\delta W_{ji} \nonumber \\
&& \phantom{\delta\Omega}=\delta\Omega_
i+\delta\Omega_j+\delta V_{ij}
+\delta V_{ji}~~;
\end{lefteqnarray}
which is the counterpart of
Eq.\,(\ref{eq:dOWVpq}), with
respect to tensor traces.

\section{Physical interpretation}
\label{phin}

In general, the virial theorem holds for potential and kinetic energies which
are averaged over a sufficiently long time  (e.g., Landau and Lifchitz 1966,
Chap.\,II, \S10; Caimmi 2007).   Similarly, the tensor virial theorem holds
for potential-energy and kinetic-energy tensor components which are averaged
over a sufficiently long time.   For sake of brevity, averaged values,
$<\Omega_u>$, $<W_{uv}>$, $<V_{uv}>$, $<T_u>$, shall be denoted as $\Omega_u$,
$W_{uv}$, $V_{uv}$, $T_u$, including related tensor components.

Aiming to a physical interpretation of potential energies and potential-energy
tensors, let an isolated subsystem, $u$, be first considered.   Accordingly,
the condition of virial equilibrium reads:
\begin{lefteqnarray}
\label{eq:veq1u}
&& \Omega_u+2T_u=0~~;
\end{lefteqnarray}
and the total energy is:
\begin{lefteqnarray}
\label{eq:E1u}
&& E_u=\Omega_u+T_u=-T_u=\frac12\Omega_u~~;
\end{lefteqnarray}
in absence of tidal interaction.

If the subsystem is infinitely dispersed i.e. each particle is infinitely
distant from each other, related energy changes are:
\begin{lefteqnarray}
\label{eq:DOm1}
&& \Delta\Omega_u=\Omega_u^\prime-\Omega_u=-\Omega_u~~; \\
\label{eq:DE1}
&& \Delta E_u=E_u^\prime-E_u=T_u-(\Omega_u+T_u)=-\Omega_u~~;
\end{lefteqnarray}
provided the kinetic energy is left unchanged, $\Delta T_u=T_u^\prime-T_u=0$,
where the prime denotes the final configuration.

Then the amount of work which must be done upon the subsystem in order to
effect the above mentioned transition is:
\begin{lefteqnarray}
\label{eq:L1u}
&& L_u=-\Delta E_u=\Omega_u~~;
\end{lefteqnarray}
where, in general, $L=-(E_{\rm F}-E_{\rm I})$ is the work required for a
transition from an initial state (energy, $E_{\rm I}$) to a final state
(energy, $E_{\rm F}$), and $L<0$ means work to be done, $L>0$ work to be
returned.   According to Eq.\,(\ref{eq:L1u}), the potential self energy,
$\Omega_u$, represents the amount of work which must be done upon the
subsystem, $u$, in order to effect an infinite dispersion of the particles
(e.g., MacMillan 1930, Chap.\,III, \S\,76).

As a second step, let two subsystems, $i$ and $j$ be considered.   The
condition of virial equilibrium for a generic subsystem, $u=i,j$, reads
(Limber 1959; Brosche et al. 1983; Caimmi et al. 1984):
\begin{lefteqnarray}
\label{eq:veq2u}
&& \Omega_u+V_{uv}+2T_u=0~~;
\end{lefteqnarray}
and the total energy is:
\begin{lefteqnarray}
\label{eq:E2u}
&& E_u=\Omega_u+W_{uv}+T_u=-T_u-Q_{uv}~~;
\end{lefteqnarray}
in presence of tidal interaction.

The kinetic energy, $T_u$, is in part macroscopic due to e.g., orbital motion
of the centre of mass and systematic rotation, and in part microscopic due to
random motions.   Systematic translation of the centre of mass is ruled out
by virial equilibrium, which implies motion of the subsystem within a limited
region of space (e.g., Landau and Lifchitz 1966, Chap.\,II, \S10; Caimmi
2007).

If the two subsystems are placed one infinitely distant from the other,
leaving both the potential self energy, $\Omega_u$, and the kinetic energy,
$T_u$, unaltered keeping the centre of mass at rest, related changes are:
\begin{lefteqnarray}
\label{eq:DOms2u}
&& \Delta^{\pprime}\Omega_u=\Omega_u^\pprime-\Omega_u=0~~; \\
\label{eq:DWs2u}
&& \Delta^{\pprime} W_{uv}=W_{uv}^\pprime-W_{uv}=-W_{uv}~~; \\
\label{eq:DTs2u}
&& \Delta^{\pprime} T_u=T_u^\pprime-T_u=0~~; \\
\label{eq:DEs2u}
&& \Delta^{\pprime} E_u=E_u^\pprime-E_u=-W_{uv}~~;
\end{lefteqnarray}
where the subsystem is no longer in virial equilibrium and must necessarily
readjust as:
\begin{lefteqnarray}
\label{eq:vep2u}
&& \Omega_u^\prime+2T_u^\prime=0~~; \\
\label{eq:Ep2u}
&& E_u^\prime=\Omega_u^\prime+T_u^\prime=-T_u^\prime=\frac12\Omega_u^\prime=
E_u^\pprime~~;
\end{lefteqnarray}
where no energy dissipation occurs.   Then related changes are:
\begin{lefteqnarray}
\label{eq:DOmp2u}
&& \Delta^{\prime}\Omega_u=\Omega_u^\prime-\Omega_u^\pprime=\Omega_u^\prime-
\Omega_u~~; \\
\label{eq:DTp2u}
&& \Delta^{\prime} T_u=T_u^\prime-T_u^\pprime=T_u^\prime-T_u~~; \\
\label{eq:DEp2u}
&& \Delta^{\prime} E_u=E_u^\prime-E_u^\pprime=0~~; \\
\label{eq:DOTu}
&& \Delta^{\prime}\Omega_u+\Delta^{\prime} T_u=0~~;
\end{lefteqnarray}
where Eq.\,(\ref{eq:DOTu}) holds via (\ref{eq:Ep2u}) and (\ref{eq:DEp2u}).

Finally, changes corresponding to the whole transition are:
\begin{lefteqnarray}
\label{eq:DOm2u}
&& \Delta\Omega_u=\Delta^\pprime\Omega_u+\Delta^{\prime}\Omega_u=
\Delta^{\prime}\Omega_u~~; \\
\label{eq:DT2u}
&& \Delta T_u=\Delta^\pprime T_u+\Delta^{\prime} T_u=\Delta^{\prime} T_u=
-\Delta^{\prime}\Omega_u=-\Delta\Omega_u~~; \\
\label{eq:DE2u}
&& \Delta E_u=\Delta^\pprime E_u+\Delta^{\prime} E_u=-W_{uv}~~;
\end{lefteqnarray}
where, on the other hand:
\begin{lefteqnarray}
\label{eq:DE2u1}
&& \Delta E_u=E_u^\prime-E_u=-T_u^\prime+T_u+Q_{uv}=-\Delta T_u+Q_{uv}=
\Delta\Omega_u+Q_{uv}~~;\qquad
\end{lefteqnarray}
and the combination of Eqs.\,(\ref{eq:DE2u}) and (\ref{eq:DE2u1}) via
(\ref{eq:Q1}) yields:
\begin{lefteqnarray}
\label{eq:DOm2u1}
&& \Delta\Omega_u=-W_{uv}-Q_{uv}=-V_{uv}~~;
\end{lefteqnarray}
in terms of the potential tidal energy.

According to Eq.\,(\ref{eq:DE2u}),
the potential interaction energy, $W_{uv}$, represents the amount of work
which must be done upon the subsystem, $u$, as a whole, in order to recede up
to an infinite distance from the subsystem, $v$, keeping the centre of mass at
rest and preserving virial
equilibrium.   In this view, the sentence (MacMillan 1930, Chap.\,III, \S76):
\begin{quotation}
``Their sum $[W=W_{ij}+W_{ji}]$ represents the exhaustion of potential energy,
due to the fact that the two bodies are non infinitely far apart.''
\end{quotation}
should be interpreted.

According to
Eq.\,(\ref{eq:DOm2u1}), the potential tidal energy, $V_{uv}$, represents the
change (regardless of the sign) in potential self energy that is necessary for
$u$ subsystem maintains
virial equilibrium in absence of $v$ subsystem, keeping the centre of mass at
rest.   For further details, an interested reader is addressed to Appendix
\ref{a:coex}, where a conceptual experiment is performed.

The above considerations can be extended to tensor components, provided the
work-tensor, $L_{pq}=-[(E_{\rm F})_{pq}-(E_{\rm I})_{pq}]$, is defined, where
$(E_{\rm K})_{pq}$ is the total energy-tensor related to the initial (K = I)
and final (K = F) state of an assigned transition, and
the trace equals the related scalar work, $L$.

In the special case of homeoidally striated ellipsoids (e.g., Caimmi and Secco
2002; Caimmi 2003),
let the subsystem, $i$, be defined by an inner ellipdoid, $0\le r\le R_i$,
and let the subsystem, $j$, be defined by an outer homeoid, $R_i\le r\le R_j$,
where $r$ is the radial coordinate and $R_i$, $R_j$, define the inner and the
outer boundary, respectively, with regard to a selected direction.   Owing to
Newton's theorem (e.g., Caimmi 2003) the resulting gravitational force exerted
on $i$ from $j$ is null i.e. the gravitational potential induced by $j$ is
constant for $0\le r\le R_i$.   Accordingly, $V_{ij}=0$ via
Eq.\,(\ref{eq:Vuvpq}) and, in addition, $W_{ij}=W_{ji}$ via
Eq.\,(\ref{eq:Ws}), $Q_{ij}=-Q_{ji}$ via Eq.\,(\ref{eq:Qa}), which by use of
Eq.\,(\ref{eq:Q1}) implies the following relations:
\begin{lefteqnarray}
\label{eq:WQs}
&& W_{ji}=W_{ij}=-Q_{ij}=Q_{ji}~~; \\
\label{eq:VWs}
&& V_{ji}=W_{ji}+Q_{ji}=W_{ij}-Q_{ij}=2W_{ij}~~;
\end{lefteqnarray}
where Eq.\,(\ref{eq:VWs}) discloses that the potential tidal energy, $V_{ji}$,
is twice the work which must be done upon the inner ellipsoid in order to
recede up to an infinite distance from the outer homeoid, according to an
earlier investigation restricted to spherical symmetry (Kondratyev 2015).

The above considerations may be extended to potential-energy tensors,
$(\Omega_u)_{pq}$, $(V_{uv})_{pq}$, $(W_{uv})_{pq}$, $(Q_{uv})_{pq}$, where
Eqs.\,(\ref{eq:veq1u})-(\ref{eq:VWs}) can be translated to related tensor
components.

\section{Tidal radius}
\label{trats}

Tidal effects do not necessarily imply stripping, in that gravitational forces
from different subsystems could exhibit a similar orientation.   For instance,
let the centre of mass of a spherical-symmetric galaxy lie within the nuclear
star cluster, and let a test particle of unit mass be located on the cluster
surface along the straight line joining the galaxy and cluster centre of mass.
It is apparent the gravitational force from the galaxy and the cluster, acting
on the above mentioned test particle, point along the same direction towards
related centre of mass (e.g., Caimmi 2015), which implies no tidal stripping
from the cluster surface.

In presence of stripping, the tidal radius of a subsystem can be defined using
either a local (i.e. involving force balance on a test particle e.g., von
Hoerner 1958; Vesperini 1997; Brosche et al. 1999; Caimmi 2015; Gajda and
Lokas 2016) or a global (i.e. involving energy balance on the whole subsystem
e.g., Caimmi and Secco 2003; Osipkov 2006) criterion.   On the other hand, in
absence of
stripping, the tidal radius of a subsystem has necessarily to be defined via a
global criterion (Secco 2000, 2001, 2005).

In the special case of similar and similarly placed spheroids, the tidal
radius for the inner component can be related to a special configuration where
the kinetic energy, $2T_i=-\Omega_i-V_{ij}$, as a function of the major
semiaxis, $a_i$, attains an extremum point (minimum) for fixed major semiaxis,
$a_j$, and masses, $M_i$, $M_j$, provided the two subsystems interact only via
gravitation and the virial theorem holds for each one (Secco 2000, 2001,
2005).   For
sufficiently steep density profiles, no extremum point occurs and no value can
be assigned to the tidal radius.   For further details, an interested reader
is addressed to the above quoted parent papers.

Aiming to a general criterion which can be applied regardless of subsystem
density profile and shape, a different attempt shall be exploited here.   Let
two subsystems interact only via gravitation and the virial theorem hold for
each one.   In the general case where the subsystems are not concentric, a
necessary condition for virial equilibrium is that related centres of mass
move along orbits within a limited region of space, which implies kinetic
energy is
partly due to systematic (orbital at least) motions and partly to random
motions.   If orbits lie outside an equipotential surface, the virial theorem
must be related to values averaged on a time, $\tau$, largely exceeding the
orbital period, $\tau_{\rm orb}$, and the notation has to be intended as
$\Phi_u=<\Phi_u>_\tau$, $\Phi=\Omega, T$; $\Psi_{uv}=<\Psi_{uv}>_\tau$,
$\Psi=W, V, Q$; $\tau\gg\tau_{\rm orb}$.
For further details, an interested reader is addressed to specific textbooks
(e.g., Landau and Lifchitz 1966, Chap.\,II, \S10).

For sake of simplicity, it shall be intended in the following that subsystem
centre of mass moves along a fictitious circular orbit where potential and
kinetic energy equal related averaged values along the real orbit.
With regard to the generic subsystem, $u$, the condition of virial
equilibrium and the total energy are expressed by Eqs.\,(\ref{eq:veq2u}) and
(\ref{eq:E2u}), respectively.

If the subsystem, $v$, is instantaneously dispersed to infinite distance, 
the remaining one, keeping the centre of mass at rest, relaxes to a virialized
configuration where the total energy, via Eqs.\,(\ref{eq:veq2u}),
(\ref{eq:E2u}), reads:
\begin{lefteqnarray}
\label{eq:Epu}
&& E_u^\prime=\Omega_u^\prime+T_u^\prime=\Omega_u+T_u=-V_{uv}-T_u~~;
\end{lefteqnarray}
and the energy change amounts to $\Delta E=E_u^\prime-E_u=-W_{uv}$,
conformly to Eq.\,(\ref{eq:DEs2u}).
Keeping the centre of mass at rest implies conversion of translation kinetic
energy into systematic either rotation or oscillation kinetic energy where the
latter, in turn, implies conversion of systematic oscillation into random
kinetic energy via violent relaxation [21].

The final state is bound or unbound according if $E_u^\prime<0$ or
$E_u^\prime>0$, respectively.   The limiting case, $E_u^\prime=0$, can be
expressed as:
\begin{lefteqnarray}
\label{eq:tiru}
&& \Omega_u=-T_u~~;\qquad V_{uv}=-T_u~~;\qquad\Omega_u=V_{uv}~~;
\end{lefteqnarray}
and the radius (intended as the distance from the centre of mass to the
boundary along a selected direction), $R_u^\ast$, for which
Eq.\,(\ref{eq:tiru}) holds, is defined as tidal radius (along that direction)
of $u$ subsystem.   While the extremum point of the kinetic energy, $T_i$, as
a function of the major semiaxis, $a_i$, implies $\Omega_i\approx V_{ij}$ in
the special case of similar and similarly placed spheroids (Secco 2000), 
$\Omega_i=V_{ij}$ in general via Eq.\,(\ref{eq:tiru}).

%
In the special case of homogeneous spheres, one completely lying within the
other, the potential self, interaction, tidal and residual energy are
expressed as:
\begin{lefteqnarray}
\label{eq:Omsf}
&& \Omega_i=-\frac35\frac{GM_i^2}{a_i}~~;\qquad \Omega_j=-\frac35\frac{GM_j^2}
{a_j}~~; \\
\label{eq:Wsf}
&& W_{ij}=-\frac35\frac{GM_i^2}{a_i}\frac m{y^3}\left(\frac54y^2-\frac14-\frac
5{12}y_0^2\right);\qquad W_{ji}=W_{ij}~~; \\
\label{eq:Vsf}
&& V_{ij}=-\frac35\frac{GM_i^2}{a_i}\frac m{y^3}\left(1+\frac53y_0^2\right);
\qquad V_{ji}=-\frac35\frac{GM_i^2}{a_i}\frac m{y^3}\left(\frac52y^2-\frac32-
\frac52y_0^2\right);\qquad \\
\label{eq:Qsf}
&& Q_{ij}=-\frac35\frac{GM_i^2}{a_i}\frac m{y^3}\left[\frac54\left(1-y^2
\right)+\frac{25}{12}y_0^2\right];\qquad Q_{ji}=-Q_{ij}; \\
\label{eq:my}
&& m=\frac{M_j}{M_i}~;\quad y=\frac{a_j}{a_i}~;\quad y\ge1~;\quad y_0=
\frac{R_0}{a_i}~;\quad0\le y_0\le y-1~;\qquad
\end{lefteqnarray}
where the indexes, $i$, $j$, label the embedded and the embedding sphere,
respectively, $M$ and $a$ denote mass and radius, respectively, and
$R_0$ is the distance between the centre of the embedding and the embedded
sphere.   For detailed calculations including potential-energy tensors, an
interested reader is addressed to Appendix \ref{a:spij}.

Accordingly, Eq.\,(\ref{eq:tiru}) via (\ref{eq:Omsf}) and (\ref{eq:Vsf})
takes the form:
\begin{lefteqnarray}
\label{eq:eq3i}
&& \frac m{y^3}\left(1+\frac53y_0^2\right)=1~~; \\
\label{eq:eq3j}
&& \frac m{y^3}\left(\frac52y^2-\frac32-\frac52y_0^2\right)=\frac{m^2}y~~;
\end{lefteqnarray}
for $i$ and $j$ subsystem, respectively.

Related tidal radii are $a_i^\ast=a_j/y_i^\ast$ and
$a_j^\ast=a_iy_j^\ast$, where $y_i^\ast$, $y_j^\ast$, are positive real
solutions of Eq.\,(\ref{eq:eq3i}), (\ref{eq:eq3j}), respectively, and $y_0$
can be expressed in terms of $y$ as:
\begin{lefteqnarray}
\label{eq:y0}
&& y_0=\zeta(y-1)~~;\qquad 0\le\zeta\le1~~;\qquad y\ge1~~;
\end{lefteqnarray}
accordingly, Eqs.\,(\ref{eq:eq3i})-(\ref{eq:eq3j}) translate into:
\begin{lefteqnarray}
\label{eq:eq3yi}
&& 3y^3-5m\zeta^2y^2+10m\zeta^2y-(5\zeta^2+3)m=0~~; \\
\label{eq:eq3yj}
&& [5(1-\zeta^2)-2m]y^2+10\zeta^2y -(5\zeta^2+3)=0~~;
\end{lefteqnarray}
in the special case of concentric spheres, $\zeta=0$, the solutions of 
Eqs.\,(\ref{eq:eq3yi}) and (\ref{eq:eq3yj}) are:
\begin{lefteqnarray}
\label{eq:eqcyi}
&& y_i^\ast=m^3~~;\qquad m\ge1~~; \\
\label{eq:eqcyj}
&& y_j^\ast=\left(\frac3{5-2m}\right)^{1/2}~~;\qquad1\le m<\frac52~~;
\end{lefteqnarray}
owing to the condition, $y\ge1$.

Turning to the general case, a third-degree equation, Eq.\,(\ref{eq:eq3yi}),
and a second-degree equation, Eq.\,(\ref{eq:eq3yj}), have to be solved for
assigned $\zeta$.  In the latter alternative, real solutions occur provided
the discriminant is nonnegative, which is equivalent to:
\begin{lefteqnarray}
\label{eq:mdel}
&& m\le\frac52\frac{3+2\zeta^2}{3+5\zeta^2}<\frac52~~;\qquad0<\zeta\le1~~;
\end{lefteqnarray}
and the solution of Eq.\,(\ref{eq:eq3yj}) reads:
\begin{lefteqnarray}
\label{eq:seqj}
&& y_j^\ast=\frac{-5\zeta^2\mp[5(3+2\zeta^2)-2m(3+5\zeta^2)]^{1/2}}
{5(1-\zeta^2)-2m}~~;
\end{lefteqnarray}
where the condition, $y\ge1$, implies the following inequality:
\begin{lefteqnarray}
\label{eq:mj}
&& 1\le m\le\frac52\frac{3+2\zeta^2}{3+5\zeta^2}<\frac52~~;
\end{lefteqnarray}
which defines the domain of the reduced mass, $m$, in the case under
discussion.   For further details, an interested reader is addressed to
Appendix \ref{a:vrsj}.
In the special case of concentric spheres, $\zeta=0$, Eq.\,(\ref{eq:mj}) reads
$1\le m<5/2$ according to Eq.\,(\ref{eq:eqcyj}).  In the special case of
tangent spheres, $\zeta=1$, Eq.\,(\ref{eq:mj}) reads $1\le m\le25/16$.

\begin{figure*}[t]
\begin{center}
\includegraphics[scale=0.8]{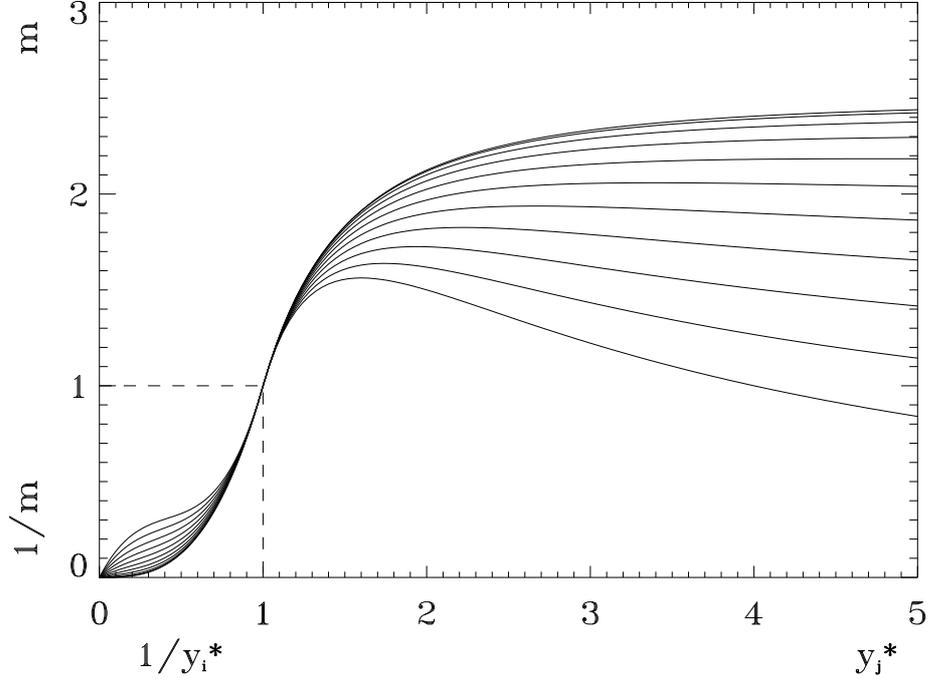}
\caption{The reduced mass, $1/m=M_i/M_j$, vs. the reduced
tidal radius, $1/y_i^\ast=a_i^\ast/a_j$, $0\le1/y_i^\ast\le1$, 
(bottom left box) and the reduced mass, $m=M_j/M_i$, vs. 
the reduced tidal radius, $y_j^\ast=a_j^\ast/a_i$, $y_j^\ast\ge1$,
for $\zeta=\ell/10$, $0\le\ell\le10$, $\ell$ integer.
Curves from top to bottom
relate to increasing $\zeta$ outside the box and to decreasing $\zeta$ inside
the box.   See text for further details.}
\label{f:tis2}
\end{center}
\end{figure*}

The reduced mass, $1/m=M_i/M_j$ and $m=M_j/M_i$, as a function of the reduced
tidal radius, $1/y_i^\ast=a_i^\ast/a_j$ and $y_j^\ast=a_j^\ast/a_i$, can be
inferred from Eq.\,(\ref{eq:eq3i}) and (\ref{eq:eq3j}), respectively, as:
\begin{lefteqnarray}
\label{eq:myti}
&& \frac1m=\frac53\zeta^2\frac1{y_i^\ast}-\frac{10}3\zeta^2\frac1
{(y_i^\ast)^2}+\left(\frac53\zeta^2+1\right)\frac1{(y_i^\ast)^3}~~; \\
\label{eq:mytj}
&& m=\frac{5(1-\zeta^2)(y_j^\ast)^2+10\zeta^2y_j^\ast-(5\zeta^2+3)}
{2(y_j^\ast)^2}~~;
\end{lefteqnarray}
where, in particular, $1/m\to0$ as $1/y_i^\ast\to0$, $1/m=1$ as
$1/y_i^\ast=1$, and $m\to5(1-\zeta^2)/2$ as $y_j^\ast\to+\infty$, $m=1$ as
$y_j^\ast=1$.   The existence of an extremum point (maximum) at
$y_j^\ast=1+(3/5)(1/\zeta^2)$ can also be ascertained, where
$m=(5/2)(3+2\zeta^2)/(3+5\zeta^2)$.   The special cases, $\zeta=\ell/10$,
$0\le\ell\le10$, $\ell$ integer, are plotted in Fig.\,\ref{f:tis2}.
An inspection of Fig.\,\ref{f:tis2} shows the occurrence of oblique
inflection points for values of $\zeta$ sufficiently close to unity i.e.
sufficiently large distance between the centre of the embedding and the
embedded sphere.   It is apparent the reduced tidal radius, $1/y_i^\ast$, can
be defined for reduced masses within the range, $0\le1/m\le1$, with regard to
the embedded spere.   Conversely, the reduced tidal radius, $y_j^\ast$, can be
defined for reduced masses within the range, $0<m<5/2$, with regard to the
embedding sphere.

\section{Discussion}\label{disc}

The above results, concerning explicit expression and physical interpretation
of potential energies and potential-energy tensors, related variations, and a
global criterion for the definition of the tidal radius, are restricted to
two-component systems for simplicity.   On the other hand, an extension can be
done to multi-component systems by (a) dealing separately with all the pairs
made of a selected subsystem and one among the others, and (b) summing up the
results due to the additivity of the gravitational potential and the tensor
potential.

The basic idea is that considering each subsystem in virial equilibrium under
the tidal action from the other(s) allows a larger amount of information  with
respect to the whole system (Limber 1959; Brosche et al. 1983; Caimmi et al.
1984; Caimmi and Secco 1992).   In this view, a physical interpretation of the
potential interaction energy and potential tidal energy can shortly be stated
as follows.   Given two subsystems, $u$ and $v$, subjected to gravitation
only, the potential interaction energy, $W_{uv}=W_{vu}$, represents the amount
of work which must be done on $u$ as a whole, in order to recede up to an
infinite distance from $v$ preserving virial equilibrium, and the potential
tidal energy, $V_{uv}=W_{uv}+Q_{uv}=W_{vu}-Q_{vu}$, represents the change
(regardless of the sign) in
potential self energy, $\Delta\Omega_u$, that is
necessary for $u$ maintains virial equilibrium in absence of $v$, in any case
keeping the centre of mass at rest.

It is widely accepted large-scale astrophysical objects are made of at least
two components, such as visible baryonic (including leptons)-dark nonbaryonic
matter, bulge-disk, bulge-halo, compact body-accretion disk, and so on.   For
concentric subsystems, a definition of tidal radius, necessarily in absence of
stripping, could be highly rewarding (e.g., Secco 2000, 2001, 2005).   To this
respect, a guidance example is restricted to homogeneous spheres for
simplicity but, on the other hand, allows a complete description of a
subsystem
completely lying within the other, where extreme situations are concentric
spheres and tangent spheres, respectively.   For instance, a description in
terms of truncated, singular isothermal spheres can be expressed analytically
only for sufficiently large distance between the centre of the embedding and
the embedded sphere (Caimmi and Secco 2003; Caimmi 2004).

The presence of a nuclear star cluster in the Galaxy (e.g., Kondratyev 2015;
Fritz et al. 2016)
and similar or less massive galaxies (e.g., Georgiev et al. 2016) invokes
a natural application of the criterion exploited in the current paper for the
definition of tidal radius, extended to globular clusters.   To this aim,
three models shall be discussed,
namely (i) homogeneous spheres; (ii) truncated, singular isothermal spheres;
in both cases, one completely lying within the other, and (iii) a
heterogeneous sphere completely lying within a Roche system i.e. a mass point
surrounded by a vanishing atmosphere.   The subsystems,
representative of a globular cluster and the Galaxy, shall be denoted as
$i={\rm C}$ and $j={\rm G}$, respectively.

\begin{figure*}[t]
\begin{center}
\includegraphics[scale=0.8]{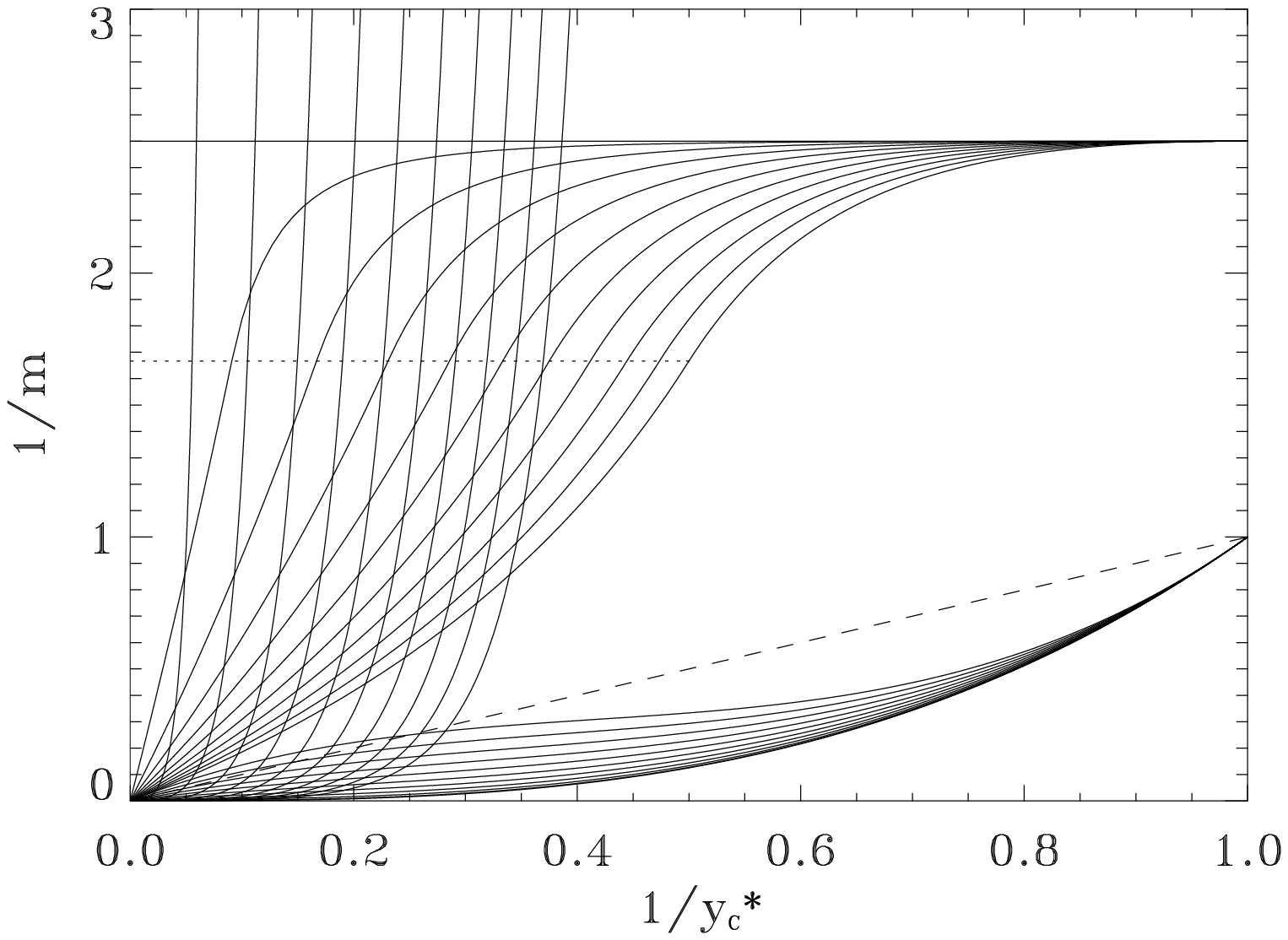}
\caption{The reduced mass, $1/m=M_{\rm C}/M_{\rm G}$, vs. the reduced
tidal radius, $1/y_{\rm C}^\ast=a_{\rm C}^\ast/a_{\rm G}$, for different
models of globular cluster within the Galaxy: (i) homogeneous spheres one
completely lying within the other [from (0,0) to (1,1)]; (ii) truncated,
singular isothermal spheres one completely lying within the other (dashed
line); (iii) a homogeneous sphere completely lying within a Roche system
[from (0,0) to (1,5/2); the locus of configurations where the mass point lies
on the boundary of the sphere, $1/m=5/3$, is represented by the dotted
horizontal line].
Divergent curves are determined according to (iv) von Hoerner's criterion
for the definition of tidal radius.  Different lines of the same kind relate
to $\zeta=\ell/10$, $0\le\ell\le10$, $\ell$ integer, from bottom to top [model
(i)]; from left to right [models (iii) and (iv)], where the vertical axis and
the origin, respectively, correspond
to $\zeta=0$; with no correlation, implying coincident lines [model (ii)].
See text for further details.}
\label{f:tis4}
\end{center}
\end{figure*}

Concerning homogeneous spheres, the plot of the reduced mass,
$1/m=M_{\rm C}/M_{\rm G}$, vs. the reduced tidal radius,
$1/y_{\rm C}^\ast=a_{\rm C}^\ast/a_{\rm G}$, is shown in the bottom left box
of Fig.\,\ref{f:tis2} and zoomed in Fig.\,\ref{f:tis4} $(0\le1/m\le1)$, where
the lower curve $(\zeta=0)$ represents concentric spheres i.e. the nuclear
star cluster, while higher curves $(\zeta>0)$ represent increasingly distant
globular clusters up to a tangential configuration $(\zeta=1)$ with respect to
the Galaxy.

Concerning trunceted, singular isothermal spheres, the reduced mass as a
function of the reduced tidal radius is expressed as:
\begin{lefteqnarray}
\label{eq:eqiyi}
&& \frac1m=\frac1{y_{\rm C}^\ast}~~;
\end{lefteqnarray}
which is acceptable to a good extent for $y>y_0\gg1$, or $0\ll\zeta\le1$.
For further details, an interested reader is addredded to Appendix
\ref{a:siso}.   The dependence of the reduced mass, $1/m$, on the reduced
tidal radius, $1/y_{\rm C}^\ast$, is linear and independent on $\zeta$, as
shown in Fig.\,\ref{f:tis4} (dashed line).

Concerning a heterogeneous sphere completely lying within a Roche system (mass
point surrounded by a vanishing atmosphere), the
reduced mass as a function of the reduced tidal radius is expressed as:
\begin{lefteqnarray}
\label{eq:eeiyi}
&& \frac1m=\frac1{\nu_\Omega}\frac1\zeta\frac1{y_{\rm C}^\ast}\left(1-\frac1
{y_{\rm C}^\ast}\right)^{-1}~~;\qquad0<\frac1{y_{\rm C}^\ast}\le\frac\zeta
{1+\zeta}~~;\qquad0\le\zeta\le1~~;\qquad
\end{lefteqnarray}
where $\nu_\Omega$ is a factor which depends on the density profile within
the sphere, with regard to an external mass point, and:
\begin{lefteqnarray}
\label{eq:eoiyi}
&& \frac1m=\frac53\left[\frac32-\frac12\zeta^2\left(\frac1{y_{\rm C}^\ast}
\right)^{-2}\left(1-\frac1{y_{\rm C}^\ast}\right)^2\right];~1\ge\frac1
{y_{\rm C}^\ast}>\frac\zeta{1+\zeta};~0\le\zeta\le1;\quad
\end{lefteqnarray}
in the special case of a homogeneous sphere $(\nu_\Omega=3/5)$, with regard to
an internal mass point.   The functions, expressed by Eqs.\,(\ref{eq:eeiyi})
and (\ref{eq:eoiyi}), join at $1/y_{\rm C}^\ast=\zeta/(1+\zeta)$ as $1/m=5/3$.
For further details, an interested reader is addressed to Appendix
\ref{a:hsmp}.

The dependence of the reduced mass, $1/m$, on the reduced tidal radius,
$1/y_{\rm C}^\ast$, is shown in Fig.\,\ref{f:tis4} $(0\le1/m<5/2)$ in the
special cases, $\zeta=\ell/10$, $0\le\ell\le10$, $\ell$ integer.
Configurations
where the mass point lies on the surface of the sphere are marked by a
horizontal dotted line $(1/m=1/\nu_\Omega=5/3)$.    By comparison, the trend
related to a classical criterion for the definition of tidal radius (von
Hoerner 1958) is also shown in Fig.\,\ref{f:tis4}, where the domain is
$0\le1/y_{\rm C}^\ast<\zeta/(1+\zeta)$ and $1/m\to+\infty$ as
$1/y_{\rm C}^\ast\to\zeta/(1+\zeta)$.   For further details, an interested
reader is addressed to Appendix \ref{a:hsmp}.

As a guidance example, a sample of 16 globular clusters discussed in an
earlier investigation (Brosche et al. 1999) shall be considered, with the
addition of Pal5 (Odenkirchen et al. 2002; Caimmi and Secco 2003) and the
Galactic nuclear star cluster (Kondratyev 2015; Fritz et al. 2016).   The
position of an
assigned globular cluster on the $({\sf O}\,1/y\,1/m)$ plane can be inferred
from the reduced radius, $a_{\rm C}/a_{\rm G}$, and the reduced mass,
$M_{\rm C}/M_{\rm G}$; and the predicted tidal radius is related to the
fractional Galactocentric distance, $y_0=R_0/a_{\rm C}$, or the parameter,
$\zeta=y_0/(y-1)=R_0/(a_{\rm G}-a_{\rm C})$.

Cluster radii, $a_{\rm C}$,
masses, $M_{\rm C}$, Galactocentric distances, $R_0$, taken from the above
quoted references, are listed in Table \ref{t:glob} with the addition of the
inferred $y_0$.
\begin{table}
\caption{Parameters of globular clusters studied in an earlier paper (Brosche
et al. 1999), with the addition of Pal5 (Odenkirchen et al. 2002) for
different inferred masses (Caimmi and Secco 2003) and the Galactic nuclear
star cluster, NSC (Kondratyev 2015; Fritz et al. 2016).   Column caption: 1 -
name (NGC or Pal or NSC); 2 - subsystem (A - [Fe/H] $>-$1, thick disk;
B - old halo; C - young halo); 3 - observed radius, $a_{\rm C}/$pc; 4 -
galactocentric distance, $R_0/$kpc; 5 - decimal logarithm of mass,
$\log(M_{\rm C}/m_\odot)$; 6 - fractional Galactocentric distance,
$y_0=R_0/a_{\rm C}$.}
\label{t:glob}
\begin{center}
\begin{tabular}{lcrrcr} \hline
\multicolumn{1}{c}{NGC}
&\multicolumn{1}{c}{S}
&\multicolumn{1}{c}{$\frac{a_{\rm C}}{\rm pc}$}
&\multicolumn{1}{c}{$\frac{R_0}{\rm kpc}$}
&\multicolumn{1}{c}
{$\log\frac{M_{\rm C}}{m_\odot}$}
&\multicolumn{1}{c}{$y_0$} \\
\noalign{\smallskip}
     &   &                 &                &                &       \\
0104 & A &  50.7           &  7.4           & 6.16           &  146  \\ 
0362 & C &  35.7           &  9.3           & 5.75           &  261  \\ 
4147 & C &  34.5           & 21.3           & 4.85           &  617  \\
5024 & C & 119.3           & 18.8           & 5.91           &  158  \\
5272 & C & 103.0           & 12.2           & 5.95           &  118  \\ 
5466 & C & 101.4           & 17.2           & 5.23           &  170  \\
5904 & C &  63.0           &  6.2           & 5.91           &   98  \\ 
6205 & B &  55.4           &  8.7           & 5.81           &  157  \\ 
6218 & B &  21.6           &  4.5           & 5.32           &  208  \\ 
6254 & B &  27.0           &  4.6           & 5.38           &  170  \\ 
6341 & B &  35.0           &  9.6           & 5.67           &  274  \\ 
6779 & B &  25.0           &  9.7           & 5.34           &  388  \\
6838 & A &  10.1           &  6.7           & 4.61           &  663  \\
6934 & C &  37.5           & 14.3           & 5.39           &  381  \\
7078 & C &  65.7           & 10.4           & 6.05           &  158  \\ 
7089 & B &  71.1           & 10.4           & 6.00           &  146  \\ 
     &   &                 &                &                &       \\
Pal5 & C &  20\phantom{.2} & 18.6           & 3.78           &  930  \\
     &   &  20\phantom{.2} & 18.6           & 3.65           &  930  \\
     &   &  20\phantom{.2} & 18.6           & 3.15           &  930  \\
     &   &  20\phantom{.2} & 18.6           & 2.98           &  930  \\
     &   &                 &                &                &       \\
NSC  &   &   1\phantom{.2} &  0\phantom{.2} & 6\phantom{.01} &    0  \\
\noalign{\smallskip}
\hline                                                       
\end{tabular}                                                
\end{center}                                                 
\end{table}                                                  
With regard to von Hoerner's criterion, an assigned cluster is expected to
show tidal effects according if
$1/y>1/y_{\rm C}^\ast$ or $a_{\rm C}/a_{\rm G}>a_{\rm C}^\ast/a_{\rm G}$, with
the exception of the nuclear stellar cluster (NSC), where the gravitational
force from the cluster acts in the same sense as the gravitational force from
the Galaxy.

For assigned cluster parameters, the position on the $({\sf O}\,1/y\,1/m)$
plane, or its logarithmic counterpart, $[{\sf O}\log(1/y)\log(1/m)]$, depends
on the Galaxy mass, $M_{\rm G}$, and radius, $a_{\rm G}$.   More specifically,
increasing/decreasing $M_{\rm G}$ makes an assigned point shift
downwards/upwards and increasing/decreasing $a_{\rm G}$ makes an assigned
point shift leftwards/rightwards.   As an exercize, the following values have
been considered: $(M_{\rm G}/10^{10}m_\odot,a_{\rm G}/{\rm kpc})=(5,25)$,
(5,125), (50,125), hereafter quoted as case a, b, c, respectively.

\begin{figure*}[t]
\begin{center}
\includegraphics[scale=0.8]{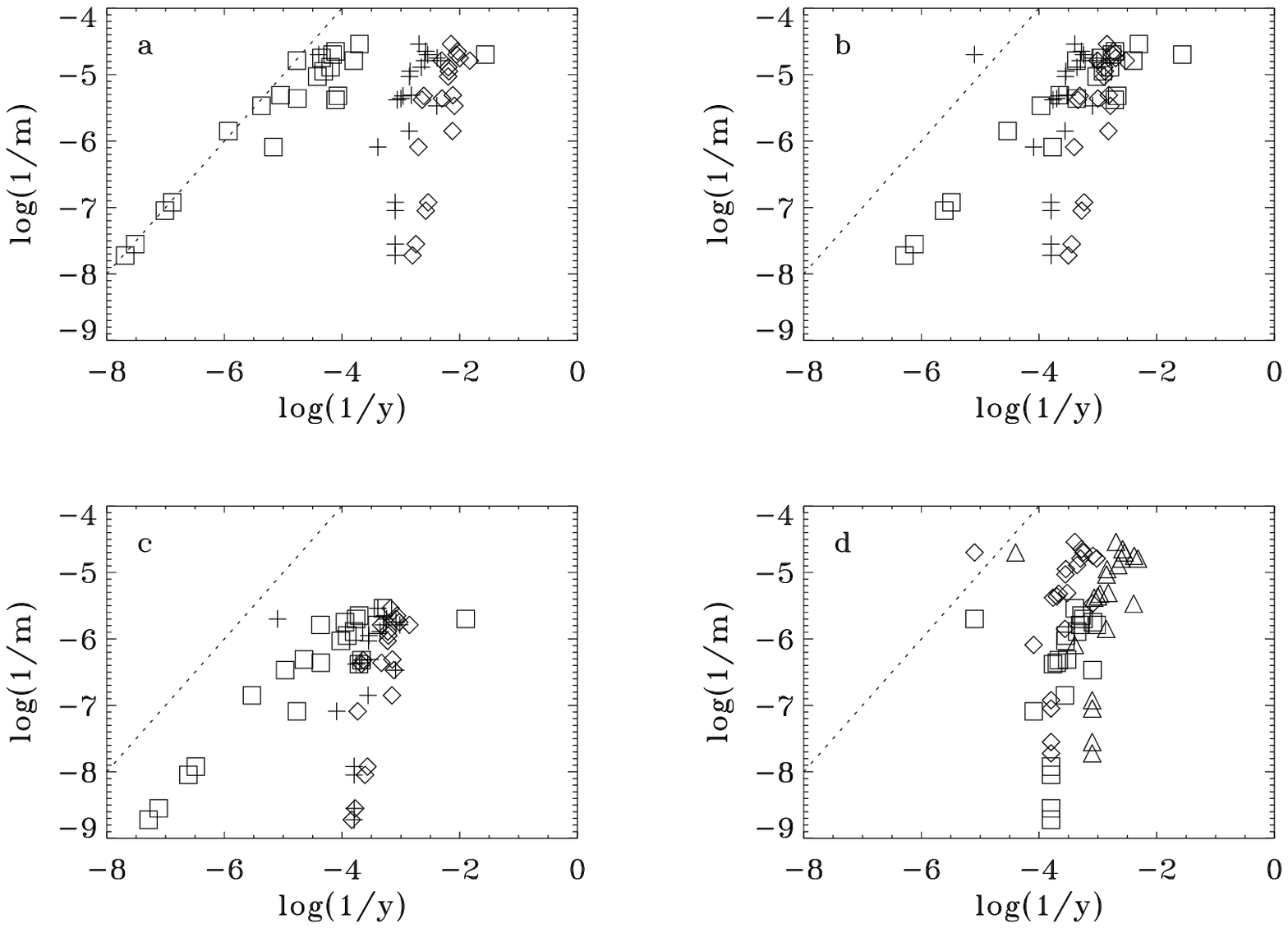}
\caption{Location of globular clusters listed in Table \ref{t:glob} on the
$[{\sf O}\log(1/y)\log(1/m)]$ plane, for assumed Galaxy mass and radius,
$(M_{\rm G}/10^{10}m_\odot, a_{\rm G}/{\rm kpc})$, equal to (5, 25), (5, 125),
(50, 125), panels a, b, c, respectively (crosses).   The whole set of
locations is collected in panel d as triangles, diamonds, squares,
corresponding to crosses on panels a, b, c, respectively.   Configurations
related to tidal radii inferred from a global criterion involving homogeneous
spheres are shown as squares, and their counterparts inferred from a classical
local criterion (von Hoerner 1958) are shown as diamonds, in both cases
concerning panels a-c.   The location of Pal15, related to four different
inferred masses, is in the lower part of each panel, $\log(1/m)\appleq-7$. The
location of the nuclear star cluster (NSC) is
near the the straight line of unit slope passing through the origin (dotted),
as can be seen in panel d.   The NSC tidal radius inferred from the global
criterion is marked by the last square on the right  in panels a-c, while the
local criterion cannot be applied in this case.  See text for further details.
Warning: symbol caption in panels a-c is different with respect to panel d, as
exlained above.}
\label{f:glog}
\end{center}
\end{figure*}

The location of globular clusters listed in Table \ref{t:glob} on the
$[{\sf O}\log(1/y)\log(1/m)]$ plane for cases a-c is shown as crosses in
Fig.\,\ref{f:glog}, panels a-c, respectively.
The whole set of locations is shown in panel d as triangles, diamonds,
squares, corresponding to crosses on panels a, b, c, respectively.
Configurations related to tidal radii inferred from a global criterion
involving homogeneous spheres, Eq.\,(\ref{eq:myti}), are shown as squares, and
their counterparts inferred from a classical local criterion (von Hoerner
1958) are shown as diamonds, in both cases concerning panels a-c.

The
location of Pal5, related to four different inferred masses, is in the lower
part of each panel, $\log(1/m)\appleq-7$.   The location of NSC is near the
straight line of unit slope passing through the origin (dotted), as clearly
shown in panel d.   The NSC tidal radius, inferred from the global criterion,
is marked by the square on the extreme right in panels a-c, while the local
criterion cannot be applied as $y_0=R_0/a_{\rm C}=0$.   NSC values listed in
Table \ref{t:glob} (Kondratyev 2015) are lower with respect to a subsequent
investigation (Fritz et al. 2016) by a factor of about 10, which makes related
points on the $[{\sf O}\log(1/y)\log(1/m)]$ plane shift upwards and rightwards
along the straight line of unit slope.

With regard to the local criterion, the Galaxy is modelled as a mass point
(von Hoerner 1958) as shown in Appendix \ref{a:VHtr}, and the results are
independent of the Galaxy radius, $a_{\rm G}$.   Accordingly, the distance
between crosses and diamonds placed on a horizontal line in panels a, b,
(where $a_{\rm G}$ attains different values) remains unchanged i.e. related
points are rigidly shifted.   The contrary holds for panel c (where
$M_{\rm G}$ attains a different value).   More specifically, cluster radius
does not exceed related tidal radius in panels a, b, while the contrary holds
in panel c for NGC 5904 and Pal5 (lowest inferred mass) and cluster radius is
slightly lower than related tidal radius for NGC 5272, NGC 5466, NGC 6254, and
Pal5 (intermediate inferred mass).

The last case (panel c) seems to be more
realistic in that (i) the mass contribution from the nonbaryonic dark halo is
included, and (ii) NGC 5466 and NGC 5904 show tidal effects (Leon et al. 2000;
Grillmair and Johnson 2006) with the addition of Pal5 even if, in this case,
due to tidal shoks during disk passages (Odenkirchen et al. 2002; Dehnen et
al. 2004).

With regard to the global criterion, the Galaxy and the cluster are modelled
as homogeneous spheres as shown in Appendix \ref{a:spij}, and the results
depend on both the Galaxy radius, $a_{\rm G}$, and mass, $M_{\rm G}$.   In
particular, $y_0\appgeq100$ from Table \ref{t:glob}, which implies
$1/m\approx(5/3)\zeta^2(1/y_{\rm C}^\ast)$ via Eq.\,(\ref{eq:myti}), where
the remaining terms on the right-hand side can be neglected to a good extent,
leaving aside NSC for which $y_0=0$, $\zeta=0$, and
$1/m=(1/y_{\rm C}^\ast)^3$.

The assumption of homogeneous, spherical-symmetric matter distribution for the
Galaxy implies tidal effects
are strongly dependent on the Galactocentric distance.   For assigned
$M_{\rm G}$ (cases a, b), the cluster radius exceeds related tidal radius
provided $R_0>4\,{\rm kpc}$ in case a and $R_0>14\,{\rm kpc}$ in case b.   For
assigned  $a_{\rm G}$ (cases b, c), the cluster radius exceeds related tidal
radius provided $R_0>14\,{\rm kpc}$ in case b and $R_0>8\,{\rm kpc}$,
$M_{\rm C}>10^5\,m_\odot$, in case c.   Leaving aside NSC,
$a_{\rm C}<a_{\rm C}^\ast$ holds for $n=0$ sample clusters in case a; $n=12$
in case b; $n=2$ in case c.

If the Galaxy and globular clusters are modelled as truncated, singular
isothermal spheres, then squares on panels a-c of Fig.\,\ref{f:glog} would
place along the straight line of unit slope passing through the origin
(dotted) via Eq.\,(\ref{eq:eqiyi}), with the exception of NSC for which
$y_0=0$ and Eq.\,(\ref{eq:eqiyi}) does not hold.   Accordingly, 
$a_{\rm C}>a_{\rm C}^\ast$ in any case.

If the Galaxy is modelled as a Roche system and globular clusters as 
homogeneous spheres, then squares on panels a-c of Fig.\,\ref{f:glog} would
place above the straight line of unit slope passing through the origin
(dotted) via Eqs.\,(\ref{eq:eeiyi})-(\ref{eq:eoiyi}).   Accordingly,
$a_{\rm C}>a_{\rm C}^\ast$ {\it a fortiori} in any case.

The global criterion, formulated in Section \ref{trats} and used in the
current application, cannot predict the occurrence of tidal effects such as
the presence of streams and tails.   On the other hand, it could provide
useful indications on the binding energy of globular clusters within the
Galaxy provided more realistic density profiles are considered.   In this
view, ``bound'' globular clusters would survive (conceptual) sudden
disappearence of the Galaxy, while ``unbound'' globular clusters would not.

\section{Conclusion} \label{conc}

Galaxies and galaxy clusters are predicted (via cosmological simulation) or
inferred (via data collection) to be made of at least two subsystems (dark
nonbaryonic and visible baryonic including leptons) which link only through
gravitation, where each component is distorted by tides from the other.   In
addition, galaxies exhibit bulge-halo and/or bulge-disk structure, where tidal
effects even in absence of accretion or merging are a common feature, in that
isolated galaxies are an exception rather than a rule.

With these ideas in mind and restricting to two-component systems, attention
has been focused on general properties of potential energies and
potential-energy tensors, including related first variations and physical
interpretation.   In addition, a global criterion for the definition of tidal
radius has been proposed and a few guidance examples, restricted to special
density profiles, have been shown.

An application has been made to a sample
of globular clusters within the Galaxy, considered in an earlier investigation
(Brosche et al. 1999), with the addition of Pal5 (Odenkirchen et al. 2002) for
different inferred masses (Caimmi and Secco 2003) and the Galactic nuclear
star cluster (Kondratyev 2015; Fritz et al. 2016).    In particular, the
extent
to which the above mentioned global criterion could provide useful indications
on the binding energy of globular clusters, has been analysed by comparison
with the results from a classical local criterion (von Hoerner 1958).

The main results of the current paper may be summarized as follows.
\begin{description}
\item[\rm{(1)}] An explicit expression has been determined for the first
variations of subsystem potential energies and potential-energy tensors, which
could be useful for e.g., practical use of virial equations in linearized form
for the treatment of the stability of a configuration (C69, Chap.\,3, \S23),
the effect of viscous dissipation on the stability (C69, Chap.\,5, \S37;
Chap.\,8, \S59), the determination of bifurcation points (C69, Chap.\,6, \S45)
and loci of neutral points belonging to third armonics (C69, Chap.\,7, \S50).
\item[\rm{(2)}] A physical interpretation has been proposed for the potential
interaction and potential tidal energy, in addition to the well known
interpretation of the potential self energy (e.g., MacMillan 1930, Chap.\,III,
\S76).   More specifically, the potential interaction energy, $W_{uv}$,
represents the amount of work which must be done upon the subsystem, $u$, as a
whole, in order to recede up to an infinite distance from the subsystem, $v$,
preserving virial equilibrium.   On the other hand, the potential tidal
energy, $V_{uv}$, represents the change (regardless of the sign) in potential
self energy, $\Omega_u$,
that is necessary for $u$ subsystem maintains virial equilibrium in absence of
$v$ subsystem.
\item[\rm{(3)}] A global criterion for the definition of subsystem tidal
radius has been inferred by requiring null total energy for $u$ subsystem in
absence of $v$ subsystem, which implies the potential self energy equals the
potential tidal energy, $\Omega_u=V_{uv}$, regardless of density profile and
slope.
\item[\rm{(4)}] Restricting to spherical-symmetric mass distributions, one
completely lying within the other, the dependence of the reduced mass,
$1/m=M_i/M_j$, on the reduced tidal radius of the embedded sphere,
$1/y_i^\ast=a_i^\ast/a_j$, for assigned fractional distance between the centre
of the
embedded and the embedding sphere, $y_0=\zeta(y-1)$, has been determined for a
few special density profiles, namely (a) both homogeneous; (b) both
truncated, singular isothermal; (c) homogeneous and Roche system i.e. mass
point surrounded by a vanishing atmosphere.  Related trends have been compared
with their counterparts inferred from a classical local criterion for the
definition of subsystem tidal radius (von Hoerner 1958).
\item[\rm{(5)}] An application has been made to a sample of Galactic globular
clusters with the addition of the Galactic nuclear star cluster (NSC), for
different values of Galaxy mass and radius.   In the more realistic case,
$(M_{\rm G}/10^{10}m_\odot,\,a_{\rm G}/{\rm kpc})=(50,125)$,
$a_{\rm C}>a_{\rm C}^\ast$, according to the local criterion, for two sample
clusters which also show tidal effects, and $a_{\rm C}<a_{\rm C}^\ast$,
according to the global criterion, for two sample clusters and NSC.
\end{description}


\begin{appendix}
\section*{Appendix}
\section{A conceptual experiment} \label{a:coex}

Further insight on the physical interpretation of the potential tidal energy
and the potential interaction energy can be gained via the following
conceptual experiment.   Let $i$, $j$, be subsystems in virial equilibrium,
under the action of gravitation only.   Let $u=i,j$, be a generic subsystem
and $v=j,i$, the remaining one.   Accordingly, the condition of virial
equilibrium and the total energy  for $u$ subsystem are expressed by
Eqs.\,(\ref{eq:veq2u}) and (\ref{eq:E2u}), respectively.   Let the following
processes take place with regard to $u$ subsystem.

First, particles are instantaneously halted and their kinetic energy is
converted into potential energy, via infinitely compressible and perfectly
elastic springs say, one per particle, which implies particles are at rest
with respect to the cosmic background radiation, say.   Second, particles are
instantaneously connected, one with the remaining others, via rigid, massless,
undeformable, infinitely thin rods to ensure potential self energy
conservation.  Third, the potential energy stored into compressed springs is
converted into translation kinetic energy of the centre of mass, which makes
the subsystem, $u$, move rigidly along a straight line at velocity,
$(2T_u/M_u)^{1/2}$.   Fourth, the centre of mass is instantaneously halted by
storing again the kinetic energy, $T_u$, into compressed springs as before
starting the translation.   Fifth, particles
are instantaneously disconnected.   Sixth, particles are instantaneously
restored free via conversion of potential energy within
springs into kinetic energy, keeping the centre of mass at rest.   Seventh,
particles are relaxed owing to the
absence of $v$ subsystem.   Eighth, particle are virialized attaining a new
equilibrium configuration.

The above mentioned states are summarized in Table \ref{t:coe}, where the
following quantities are listed:
\begin{table*}
\caption[par]{Sequential states of $u$ subsystem during the conceptual
experiment discussed in the
text, related transition time $(\Delta t)$, potential self energy (PSE), 
potential tidal energy (PTE), potential interaction energy (PIE), kinetic
energy (KE), and particle status.   The initial state, by definition, relates
to a null transition time.   Sequential states of $v$ subsystem can be
inferred by replacing the index, $u$, with the index, $v$, and vice versa.}
\label{t:coe}
\begin{center}
\begin{tabular}{ccccccc}
\hline
\multicolumn{1}{c}{state} &
\multicolumn{1}{c}{$\Delta t$} &
\multicolumn{1}{c}{PSE} &
\multicolumn{1}{c}{PTE} &
\multicolumn{1}{c}{PIE} &
\multicolumn{1}{c}{KE} &
\multicolumn{1}{c}{status} \\
\hline\noalign{\smallskip}
1 & 0         & $\Omega_u$        & $V_{uv}$ & $W_{uv}$ & $T_u$        & virialized   \\
2 & 0         & $\Omega_u$        & $V_{uv}$ & $W_{uv}$ & 0            & halted       \\
3 & 0         & $\Omega_u$        & $V_{uv}$ & $W_{uv}$ & 0            & connected    \\
4 & 0         & $\Omega_u$        & $V_{uv}$ & $W_{uv}$ & $T_u$        & started      \\
5 & $ \infty$ & $\Omega_u$        & 0        & 0        & $T_u$        & translated   \\
6 & 0         & $\Omega_u$        & 0        & 0        & 0            & halted       \\
7 & 0         & $\Omega_u$        & 0        & 0        & 0            & disconnected \\
8 & 0         & $\Omega_u$        & 0        & 0        & $T_u $       & restored      \\
9 & $\tau$    & $\Omega_u^\prime$ & 0        & 0        & $T_u^\prime$ & virialized   \\
\noalign{\smallskip}
\hline
\end{tabular}
\end{center}
\end{table*}
the transition time $(\Delta t)$, the potential self energy (PSE), the
potential tidal energy (PTE), the potential interaction energy (PIE), the
kinetic energy (KE), and the particle status.   All the transitions are
conceived as instantaneous $(\Delta t=0)$ that is true, by definition, for
the initial state, with the exception of the translation to infinite distance,
which needs an infinite time, and the last virialization, which is completed
in a relaxation time, $\tau$.

The kinetic energy is null in the state 2, 3,
6, 7, but an equivalent amount is stored as potential energy into compressed
springs, according to the above considerations.   The kinetic energy is partly
due to systematic motions and partly to random motions in the state 1,
entirely due to systematic motions in the state 4, 5, partly due to systematic
motions and partly due to random motions in the state 8, 9.

In particular,
the orbital kinetic energy in the state 1 cannot be converted into
translation kinetic energy in the state 8 to ensure subsystem confinement
within a limited region of space, implying virial equilibrium (e.g., Landau
and Lifchitz 1966, Chap.\,II, \S10; Caimmi 2007).   More specifically, the
orbital kinetic energy can be preserved in macroscopic form via conversion
into systematic rotation, or initially preserved in macroscopic form via
conversion into radial oscillations and progressively turned into microscopic
form via violent relaxation (Lynden-Bell 1967).

The transition 1-4 violates the second principle of thermodynamics in that
kinetic energy is transferred from random motions to translation motions via
counterparts of Maxwell's daemons, who are able to halt (via infinitely
compressible and perfectly elastic springs) and
connect (via rigid, massless, undeformable, infinitely thin rods) particles.
On the other hand, the total energy is left unchanged and
Eqs.\,(\ref{eq:veq2u})-(\ref{eq:E2u}) hold.

The reverse occurs for the
transition 5-8, where daemons act to transfer kinetic energy from translation
motions to random motions, conformly to the second principle of
thermodynamics, leaving the total energy unchanged even if intrinsically
different with
respect to the transition 1-4.   Accordingly, $E_u^\pprime=\Omega_u+T_u$ via
Eqs.\,(\ref{eq:E2u}) and (\ref{eq:DEs2u}).   The
subsystem, $u$, virializes through the transition 8-9, and
Eq.\,(\ref{eq:vep2u})
holds.   The transition 1-8 is equivalent to the instantaneous disappearence
of $v$ subsystem keeping the centre of mass of $u$ subsystem at rest and
leaving the kinetic energy, $T_u$, unchanged, as assumed in Section
\ref{phin}.

\section{Homogeneous spheres one completely lying within the
other}  \label{a:spij}

Let $({\sf O}_ix_1x_2x_3)$ and $({\sf O}_jX_1X_2X_3)$ be Cartesian reference
frames with origin placed on the centre of the embedded and embedding sphere,
respectively, coinciding axes, $x_1$, $X_1$, and parallel axes, $x_2$, $X_2$;
$x_3$, $X_3$.   Let ${\sf P}(x_1,x_2,x_3)\equiv{\sf P}(X_1,X_2,X_3)$ be a
generic point of the embedded sphere, $r=(x_1^2+x_2^2+x_3^2)^{1/2}$ the radial
coordinate of ${\sf P}$ with respect to $({\sf O}_ix_1x_2x_3)$, and 
$R=(X_1^2+X_2^2+X_3^2)^{1/2}$, $R_0=(X_{01}^2+X_{02}^2+X_{03}^2)^{1/2}$ the
radial coordinates of ${\sf P}$, ${\sf O}_i$, with respect to
$({\sf O}_jX_1X_2X_3)$.

Accordingly, Cartesian coordinates are related as:
\begin{lefteqnarray}
\label{eq:Xsxs}
&& X_s=x_s+\delta_{1s}R_0~~;\qquad s=1,2,3~~;
\end{lefteqnarray}
where $\delta_{pq}$ is the Kronecker symbol, and radial coordinates are
related as:
\begin{lefteqnarray}
\label{eq:RR0r}
&& R^2=R_0^2+2R_0x_1+r^2~~;
\end{lefteqnarray}
for the reference frames under consideration.

The (gravitational) tensor potential and potential within a homogeneous sphere
can be determined in a twofold manner, from (a) the general expression for
homogeneous ellipsoids (e.g., Caimmi and Secco 1992) in the spherical limit,
or (b) the general
expression for heterogeneous spheres (e.g., Caimmi and Secco 2003) in the
homogeneous limit.   With regard to the embedding sphere, the result is:
\begin{lefteqnarray}
\label{eq:Vjpq}
&& ({\cal V}_j)_{pq}(R)=\frac{GM_j}{a_j^3}\left[\frac25X_pX_q+\delta_{pq}
\left(\frac12a_j^2-\frac3{10}R^2\right)\right]~~; \\
\label{eq:Vj}
&& {\cal V}_j(R)=\frac{GM_j}{a_j^3}\left(\frac32a_j^2-\frac12R^2\right)~~;
\end{lefteqnarray}
and, in addition:
\begin{lefteqnarray}
\label{eq:DVjXp}
&& \frac{\partial{\cal V}_j}{\partial X_q}X_p=\frac{\partial{\cal V}_j}
{\partial R}
\frac{\partial R}{\partial X_q}X_p=\frac{\partial{\cal V}_j}{\partial R}\frac
{X_q}RX_p=-\frac{GM_j}{a_j^3}X_qX_p~~; \\
\label{eq:sDVjXp}
&& \sum_{s=1}^3\frac{\partial{\cal V}_j}{\partial X_s}X_s=-\frac{GM_j}{a_j^3}
\sum_{s=1}^3X_s^2=-\frac{GM_j}{a_j^3}R^2~~;
\end{lefteqnarray}
where the reference frame is $({\sf O}_jX_1X_2X_3)$.

The substitution of Eqs.\,(\ref{eq:Xsxs})-(\ref{eq:RR0r}) into
(\ref{eq:Vjpq})-(\ref{eq:sDVjXp}) after some algebra yields:
\begin{lefteqnarray}
\label{eq:Vjpq2}
&& ({\cal V}_j)_{pq}(R)=\frac{GM_j}{a_j^3}\left[\frac25\left(x_px_q+
\delta_{1p}R_0x_q+\delta_{1q}R_0x_p+\delta_{1p}\delta_{1q}R_0^2\right)\right.
\nonumber \\
&& \phantom{({\cal V}_j)_{pq}(R)=\frac{GM_j}{a_j^3}\left[\right.}+\left.
\frac12\delta_{pq}a_j^2-\frac3{10}\delta_{pq}\left(r^2+2R_0x_1+R_0^2\right)
\right]~~; \\
\label{eq:Vj2}
&& {\cal V}_j(R)=\frac{GM_j}{a_j^3}\left[\frac32a_j^2-\frac12\left(r^2+
2R_0x_1+R_0^2\right)\right]~~; \\
\label{eq:DVj2}
&& \frac{\partial{\cal V}_j}{\partial X_q}X_p=-\frac{GM_j}{a_j^3}\left(
x_qx_p+\delta_{1q}R_0x_p+\delta_{1p}R_0x_q+\delta_{1q}\delta_{1p}R_0^2\right)
~~; \\
\label{eq:sDVj2}
&& \sum_{s=1}^3\frac{\partial{\cal V}_j}{\partial X_s}X_s=-\frac{GM_j}{a_j^3}
(r^2+2R_0x_1+R_0^2)~~;
\end{lefteqnarray}
where the reference frame is $({\sf O}_ix_1x_2x_3)$.

The potential-energy tensors, $(W_{ij})_{pq}$ and $(V_{ij})_{pq}$, can be
determined from the substitution of Eq.\,(\ref{eq:Vjpq2}), (\ref{eq:DVj2}),
into (\ref{eq:Wuvpq}), (\ref{eq:Vuvpq}), respectively, and related
integration on the volume of the embedded sphere, $S_i=(4\pi/3)a_i^3$.   In
spherical coordinates, the infinitesimal volume element reads:
\begin{lefteqnarray}
\label{eq:d3Si}
&& \diff^3S_i=\diff x_1\diff x_2\diff x_3=r^2\sin\theta\diff r\diff\theta\diff
\phi~~; \\
\label{eq:sfc}
&& x_1=r\sin\theta\cos\phi~~;\qquad x_2=r\sin\theta\sin\phi~~;\qquad x_3=r
\cos\theta~~;\qquad \\
\label{eq:dsc}
&& 0\le\phi\le2\pi~~;\qquad0\le\theta\le\pi~~;\qquad0\le r\le a_1~~;
\end{lefteqnarray}
and the following relations hold:
\begin{lefteqnarray}
\label{eq:ixpxq}
&& \int_{S_i}x_px_q\diff^3S_i=\delta_{pq}\frac15S_ia_i^2~~;\qquad p=1,2,3~~;
\qquad q=1,2,3~~; \\
\label{eq:ixs}
&& \int_{S_i}x_s\diff^3S_i=0~~;\qquad s=1,2,3~~;
\end{lefteqnarray}
after transformation of Cartesian into spherical coordinates (e.g., Spiegel
1968, Chap.\,22, \S\S22.81-83).

Accordingly, the integration of the right-hand side of Eqs.\,(\ref{eq:Wuvpq})
and (\ref{eq:Vuvpq}) via (\ref{eq:Vjpq2})-(\ref{eq:ixs}) yields:
\begin{lefteqnarray}
\label{eq:Wth}
&& (W_{ij})_{pq}=-\frac15\delta_{pq}\frac{GM_i^2}{a_i}\frac m{y^3}\left[\frac
54y^2-\frac14+\left(\delta_{1p}-\frac34\right)y_0^2\right];\,
(W_{ji})_{pq}=(W_{ij})_{pq}\,;\qquad \\
\label{eq:Vth}
&& (V_{ij})_{pq}=-\frac15\delta_{pq}\frac{GM_i^2}{a_i}\frac m{y^3}\left(1+5
\delta_{1p}y_0^2\right);
\end{lefteqnarray}
and Eq.\,(\ref{eq:Quvpq}) takes the explicit form:
\begin{lefteqnarray}
\label{eq:Qth}
&& (Q_{ij})_{pq}=-\frac15\delta_{pq}\frac{GM_i^2}{a_i}\frac m{y^3}\left[\frac
54\left(1-y^2\right)+\left(4\delta_{1p}+\frac34\right)y_0^2\right];~
\nonumber \\
&& (Q_{ji})_{pq}=-(Q_{ij})_{pq}~~;\quad
\end{lefteqnarray}
finally, using Eqs.\,(\ref{eq:Quvpq}), (\ref{eq:Wst}), (\ref{eq:Qat}), yields:
\begin{lefteqnarray}
\label{eq:Vgh}
&& (V_{ji})_{pq}=(W_{ji})_{pq}+(Q_{ji})_{pq}=(W_{ij})_{pq}-(Q_{ij})_{pq}~~;
\end{lefteqnarray}
and the substitution of Eqs.\,(\ref{eq:Wth}) and (\ref{eq:Qth}) into
(\ref{eq:Vgh}) produces:
\begin{lefteqnarray}
\label{eq:Vuh}
&& (V_{ji})_{pq}=-\frac15\delta_{pq}\frac{GM_i^2}{a_i}\frac m{y^3}\left[\frac
52y^2-\frac32-\frac32\left(2\delta_{1p}+1\right)y_0^2\right]~~;
\end{lefteqnarray}
which completes the determination of potential-energy tensors in the case
under discussion.   Related traces are expressed by
Eqs.\,(\ref{eq:Wsf})-(\ref{eq:Qsf}).

\section{Reduced mass vs. reduced virial radius of the embedding sphere}
\label{a:vrsj}

With regard to the embedding and the embedded spere, the reduced radius,
$y=a_j/a_i$, by definition satisfies the inequality, $y\ge1$.   Then the
fraction on the right-hand side of Eq.\,(\ref{eq:seqj}), $y_j^\ast=N/D$,
satisfies either $N\ge D>0$ or $N\le D<0$.   An additional condition,
expressed by Eq.\,(\ref{eq:mdel}), impliyng real solutions of the
second-degree
equation, Eq.\,(\ref{eq:eq3yj}), is a nonnegative discriminant, $\Delta\ge0$.

The former alternative, $N\ge D>0$, implies the following relation:
\begin{lefteqnarray}
\label{eq:NgeD1}
&& \mp[5(3+2\zeta^2)-2m(3+5\zeta^2)]^{1/2}\ge5-2m~~;\qquad0\le\zeta\le1~~; 
\end{lefteqnarray}
where $5-2m\ge5(1-\zeta^2)-2m=D>0$, which rules out the minus on the left-hand
side.   Accordingly, Eq.\,(\ref{eq:NgeD1}) is equivalent to:
\begin{lefteqnarray}
\label{eq:NgeD2}
&& 5(3+2\zeta^2)-2m(3+5\zeta^2)\ge(5-2m)^2~~; 
\end{lefteqnarray}
which can be ordered in $m$ as:
\begin{lefteqnarray}
\label{eq:NgeD3}
&& 2m^2-(7-5\zeta^2)m+5(1-\zeta^2)\le0~~; 
\end{lefteqnarray}
where the solutions of the associated equation are:
\begin{lefteqnarray}
\label{eq:seND1}
&& m_1=1~~;\qquad m_2=\frac52(1-\zeta^2)~~; 
\end{lefteqnarray}
and the solution of the disequation reads:
\begin{lefteqnarray}
\label{eq:NgeDs}
&& \min\left[1,\frac52(1-\zeta^2)\right]\le m\le\max
\left[1,\frac52(1-\zeta^2)\right]~~;
\end{lefteqnarray}
on the other hand the condition, $D>0$, is equivalent to:
\begin{lefteqnarray}
\label{eq:Dgt0}
&& m<\frac52(1-\zeta^2)~~;
\end{lefteqnarray}
and the combination of Eqs.\,(\ref{eq:NgeDs}) and (\ref{eq:Dgt0}) yields:
\begin{lefteqnarray}
\label{eq:NDg0s}
&& 1\le m<\frac52(1-\zeta^2)~~;\qquad0\le\zeta<\sqrt{\frac35}~~;
\end{lefteqnarray}
which is the domain of reduced mass, $m=M_j/M_i$, related to the reduced tidal
radius, $y_j^\ast=a_j^\ast/a_i$, in the case under discussion.

The latter alternative, $N\le D<0$, implies the following relation:
\begin{lefteqnarray}
\label{eq:NleD1}
&& \mp[5(3+2\zeta^2)-2m(3+5\zeta^2)]^{1/2}\le5-2m~~;\qquad0\le\zeta\le1~~; 
\end{lefteqnarray}
where $5-2m>0$, owing to Eq.\,(\ref{eq:mdel}).   Accordingly, the minus on the
left-hand side of Eq.\,(\ref{eq:NleD1}) can be erased in that the remaining
inequality implies the validity of both.

Following a similar procedure as in the former case, the solution of the
disequation reads:
\begin{lefteqnarray}
\label{eq:NleDs}
&& m\le\min\left[1,\frac52(1-\zeta^2)\right]~~;\qquad m\ge\max
\left[1,\frac52(1-\zeta^2)\right]~~;
\end{lefteqnarray}
on the other hand the condition, $D<0$, is equivalent to:
\begin{lefteqnarray}
\label{eq:Dlt0}
&& m>\frac52(1-\zeta^2)~~;
\end{lefteqnarray}
and the combination of Eqs.\,(\ref{eq:NleDs}), (\ref{eq:Dlt0}), and
(\ref{eq:mdel}) yields:
\begin{lefteqnarray}
\label{eq:NDl0s}
&& \max\left[1,\frac52(1-\zeta^2)\right]\le m\le\frac52\frac
{3+2\zeta^2}{3+5\zeta^2}<\frac52~~;\qquad0<\zeta\le1~~;
\end{lefteqnarray}
which is the domain of reduced mass, $m=M_j/M_i$, related to the reduced tidal
radius, $y_j^\ast=a_j^\ast/a_i$, in the case under discussion.

The combination of Eqs.\,(\ref{eq:NDg0s}) and (\ref{eq:NDl0s}) yields
Eq.\,(\ref{eq:mj}).

\section{Truncated, singular isothermal spheres, one completely lying within
the other} \label{a:siso}

Potential energies and potential-energy tensors of truncated, singular
isothermal spheres, one completely lying within the other, can be expressed in
simple form only if the centre of the embedded sphere is sufficiently distant
from the centre of the embedding sphere.   The result is (Caimmi and Secco
2003):
\begin{lefteqnarray}
\label{eq:Ousf}
&& \Omega_i=-\frac{GM_i^2}{a_i}~~;\qquad\Omega_j=-\frac{GM_j^2}{a_j}~~; \\
\label{eq:Wijsi}
&& W_{ij}=-\frac12\frac{GM_i^2}{a_i}\frac my\left[1-\ln\frac{y_0}y-\frac1{18}
\frac1{y_0^2}\right]~~;\qquad y>y_0\gg1~~; \\
\label{eq:Vijsi}
&& V_{ij}=W_{ij}+Q_{ij}=-\frac{GM_i^2}{a_i}\frac my~~;\qquad y>y_0\gg1~~;  \\
\label{eq:Qijsi}
&& Q_{ij}=-\frac12\frac{GM_i^2}{a_i}\frac my\left[1+\ln\frac{y_0}y+\frac1{18}
\frac1{y_0^2}\right]~~;\qquad y>y_0\gg1~~; \\
\label{eq:sWQ}
&& W_{ji}=W_{ij}~~;\qquad Q_{ji}=-Q_{ij}~~; \\
\label{eq:Vjisi}
&& V_{ji}=W_{ji}+Q_{ji}=\frac{GM_i^2}{a_i}\frac my\left[\ln\frac{y_0}y+
\frac1{18}\frac1{y_0^2}\right]~~;\qquad y>y_0\gg1~~;
\end{lefteqnarray}
accordingly, Eq.\,(\ref{eq:tiru}) takes the explicit form:
\begin{lefteqnarray}
\label{eq:myisi}
&& \frac my=1~~;\qquad y>y_0\gg1~~; \\
\label{eq:myjsi}
&& \frac my\left[\ln\frac{y_0}y+\frac1{18}
\frac1{y_0^2}\right]=-\frac{m^2}y~~;\qquad y>y_0\gg1~~;
\end{lefteqnarray}
for $i$ and $j$ subsystem, respectively.

The reduced mass, $1/m=M_i/M_j$ and $m=M_j/M_i$, in terms of the reduced tidal
radius, $1/y_i^\ast=a_i^\ast/a_j$ and $y_j^\ast=a_j^\ast/a_i$, can be inferred
from Eqs.\,(\ref{eq:myisi}) and (\ref{eq:myjsi}), respectively, via
(\ref{eq:y0}).   The result is:
\begin{lefteqnarray}
\label{eq:mtisi}
&& \frac1m=\frac1{y_i^\ast}~~;\qquad 0\ll\zeta\le1~~; \\
\label{eq:mtjsi}
&& m=-\ln\zeta-\ln\left(1-\frac1{y_j^\ast}\right)-\frac1{18}\frac1{\zeta^2}
\left(\frac1{y_j^\ast}\right)^2\left(1-\frac1{y_j^\ast}\right)^{-2};~~
0\ll\zeta\le1~;\qquad
\end{lefteqnarray}
where $m\to-\infty$ as $y_j^\ast\to1^+$, $m\to-\ln\zeta$ as
$y_j^\ast\to+\infty$,
and the absence of extremum points implies the existence of a single zero,
$y_{0,j}^\ast$, for $m$.   Accordingly, the tidal radius of the embedding
sphere can be defined within the range, $y_{j}^\ast>y_{0,j}^\ast$, keeping in
mind $y_{0,j}^\ast\to+\infty$ as $\zeta\to1^-$.

In the special case of a globular cluster $(i={\rm C})$ within the Galaxy
$(j={\rm G})$, Eq.\,(\ref{eq:mtisi}) reduces to (\ref{eq:eqiyi}).

\section{A heterogeneous sphere completely lying within a Roche system}
\label{a:hsmp}

A Roche system (mass point surrounded by a vanishing atmosphere) is preferred
to a ``naked'' mass point in that it can be conceived as a heterogeneous
sphere with infinite concentration.   Let $({\sf O}X_1X_2X_3)$ be a Cartesian
reference frame with the origin placed on the mass point and axis, $X_1$,
passing through the centre of the heterogeneous sphere.   Let
$R_0=(X_{01}^2+X_{02}^2+X_{03}^2)^{1/2}=X_{01}$ be the distance between
the centre and the mass point.   As the tidal radius cannot be
defined for a mass point, considerations shall be restricted to the
heterogeneous sphere.   Let $\overline R_0$ be the radius of a fictitious
circular orbit of the centre of the sphere around the mass
point where the virial theorem is satisfied and, in consequence, related
potential and kinetic energy equal the mean values along the real orbit.
Further attention shall be restricted to the fictitious orbit for simplicity,
hence $\overline R_0=R_0$.   In general, the mass point can be inside or
outside the heterogeneous sphere.   The two possibilities shall be discussed
separately.

\subsection{Mass point outside the heterogeneous sphere}
\label{a:mpos}

The gravitational potential of the heterogeneous sphere in ${\sf O}$ is (e.g.,
MacMillan 1930, Chap.\,II, \S29):
\begin{lefteqnarray}
\label{eq:Pse}
&& {\cal V}_i({\sf O})=\frac{GM_i}{R_0}~~;
\end{lefteqnarray}
and, in addition:
\begin{lefteqnarray}
\label{eq:dPse}
&& \left(\frac{\partial{\cal V}_i}{\partial X_s}\right)_{\sf O}=
\left(\frac{\partial{\cal V}_i}{\partial R}\frac{\partial R}{\partial X_s}
\right)_{\sf O}=-\frac{GM_i}{R_0^2}\frac{X_{0s}}{R_0}=-\delta_{1s}
\frac{GM_i}{R_0^2}~~; \\
\label{eq:dPs}
&& \sum_{s=1}^3\left(\frac{\partial{\cal V}_i}{\partial X_s}X_s
\right)_{\sf O}=0~~;
\end{lefteqnarray}
where $\delta_{pq}$ is the Kronecker symbol.

Related potential energies are:
\begin{lefteqnarray}
\label{eq:Oei}
&& \Omega_i=-\nu_\Omega\frac{GM_i^2}{a_i}~~; \\
\label{eq:Weji}
&& W_{ji}=-\frac12M_j{\cal V}_i({\sf O})=-\frac12\frac{GM_iM_j}{R_0}~~; \\
\label{eq:Veji}
&& V_{ji}=M_j\sum_{s=1}^3\left(\frac{\partial{\cal V}_i}{\partial X_s}X_s
\right)_{\sf O}=0~~; \\
\label{eq:Qeji}
&& Q_{ji}=V_{ji}-W_{ji}=\frac12\frac{GM_iM_j}{R_0}~~; \\
\label{eq:Veij}
&& V_{ij}=W_{ij}+Q_{ij}=-\frac{GM_iM_j}{R_0}~~;
\end{lefteqnarray}
where the last relation is owing to the symmetry of the potential interaction
energy via Eq.\,(\ref{eq:Ws}), $W_{ij}=W_{ji}$, and to the antisymmetry of
the potential residual energy via Eq.\,(\ref{eq:Qa}), $Q_{ij}=-Q_{ji}$.

With regard to the heterogeneous sphere, Eq.\,(\ref{eq:tiru}) via
(\ref{eq:my}), (\ref{eq:y0}), (\ref{eq:Oei}), (\ref{eq:Veij}), takes the
explicit form:
\begin{lefteqnarray}
\label{eq:rtes}
&& \nu_\Omega=\frac1\zeta\frac m{y-1}~;\quad y_0=\zeta(y-1)\ge1~;\quad
y\ge\frac{1+\zeta}\zeta~;\quad0\le\zeta\le1~;\qquad
\end{lefteqnarray}
where $\nu_\Omega$ is a factor which depends on the density profile within the
sphere e.g., $\nu_\Omega=3/5$ for the homogeneous sphere and $\nu_\Omega=1$
for the truncated, singular isothermal sphere.

The reduced mass, $1/m=M_i/M_j$, in terms of the reduced tidal radius,
$1/y_i^\ast=a_i^\ast/a_j$, can be inferred from Eq.\,(\ref{eq:rtes}) as:
\begin{lefteqnarray}
\label{eq:cuse}
&& \frac1m=\frac1{\nu_\Omega}\frac1\zeta\frac1{y_i^\ast}\left(1-\frac1
{y_i^\ast}\right)^{-1};\qquad0<\frac1{y_i^\ast}\le\frac\zeta{1+\zeta}~~;\qquad
0\le\zeta\le1~~;\qquad
\end{lefteqnarray}
where $1/m=1/\nu_\Omega$ as $1/y_i^\ast=\zeta/(1+\zeta)$.   In the special
case of a globular cluster $(i={\rm C})$ within the Galaxy $(j={\rm G})$,
Eq.\,(\ref{eq:cuse}) reduces to (\ref{eq:eeiyi}).

\subsection{Mass point inside the heterogeneous sphere}
\label{a:mpos}

The gravitational potential of the heterogeneous sphere in ${\sf O}$ is (e.g.,
Caimmi 2003):
\begin{lefteqnarray}
\label{eq:Pot}
&& {\cal V}_i({\sf O})={\cal V}_i^{\rm (int)}({\sf O})+{\cal V}_i^{\rm (ext)}
({\sf O})~~; \\
\label{eq:Poe}
&& {\cal V}_i^{\rm (ext)}({\sf O})=\frac{GM_i(R_0)}{R_0}=\frac{GM_i}{a_i}\frac
{M_i(R_0)}{M_i}\frac{a_i}{R_0}~~; \\
\label{eq:Poi}
&& {\cal V}_i^{\rm (int)}({\sf O})=2\pi G\rho_{0,i}a_i^2F(\xi_{\sf O})~~; \\
\label{eq:Fc}
&& F(\xi)=2\int_\xi^1f(\xi)\xi\diff\xi~~; \\
\label{eq:rhf}
&& \rho_i(r)=\rho_{0,i}f(\xi)~~;\qquad f(0)=1~~;\qquad\xi=\frac r{a_i}~~;
\end{lefteqnarray}
where $\rho_{0,i}$ is the central density, $\rho_i(r)$ the density profile,
$M_i(r)$ the mass distribution, $\xi$ a reduced radial coordinate,
$\xi_{\sf O}=R_0/a_i=y_0\le1$.

From this point on, attention shall be restricted to a homogeneous sphere for
simplicity, which implies $\rho_{0,i}=3M_i/(4\pi a_i^3)$,
$M_i(R)/M_i=R^3/a_i^3$, $f(\xi)=1$, $F(\xi)=1-\xi^2$.   Accordingly, 
Eqs.\,(\ref{eq:Poe}) and (\ref{eq:Poi}) reduce to:
\begin{lefteqnarray}
\label{eq:Pohe}
&& {\cal V}_i^{\rm (ext)}({\sf O})=\frac{GM_i}{a_i}\frac{R_0^2}{a_i^2}~~; \\
\label{eq:Pohi}
&& {\cal V}_i^{\rm (int)}({\sf O})=\frac32\frac{GM_i}{a_i}\left(1-\frac{R_0^2}
{a_i^2}\right)~~;
\end{lefteqnarray}
and Eq.\,(\ref{eq:Pot}) takes the explicit form:
\begin{lefteqnarray}
\label{eq:Poh}
&& {\cal V}_i({\sf O})=\frac{GM_i}{a_i}\left(\frac32-\frac12\frac{R_0^2}
{a_i^2}\right)~~;
\end{lefteqnarray}
accordingly, related potential energies are:
\begin{lefteqnarray}
\label{eq:Whji}
&& W_{ji}=-\frac12M_j{\cal V}_i({\sf O})=-\frac12\frac{GM_iM_j}{a_i}
\left(\frac32-\frac12\frac{R_0^2}{a_i^2}\right)~~; \\
\label{eq:Vhji}
&& V_{ji}=M_j\sum_{s=1}^3\left(\frac{\partial{\cal V}_i}{\partial X_s}X_s
\right)_{\sf O}=0~~; \\
\label{eq:Qhji}
&& Q_{ji}=V_{ji}-W_{ji}=\frac12\frac{GM_iM_j}{a_i}
\left(\frac32-\frac12\frac{R_0^2}{a_i^2}\right)~~; \\
\label{eq:Vhij}
&& V_{ij}=W_{ij}+Q_{ij}=-\frac{GM_iM_j}{a_i}
\left(\frac32-\frac12\frac{R_0^2}{a_i^2}\right)~~;
\end{lefteqnarray}
where the last relation is owing to the symmetry of the potential interaction
energy via Eq.\,(\ref{eq:Ws}), $W_{ij}=W_{ji}$, and the antisymmetry of the
potential resuidual energy via Eq.\,(\ref{eq:Qa}), $Q_{ij}=-Q_{ji}$.

With regard to the homogeneous sphere, Eq.\,(\ref{eq:tiru}) via (\ref{eq:my}),
(\ref{eq:y0}), (\ref{eq:Oei}), (\ref{eq:Vhij}), takes the explicit form:
\begin{lefteqnarray}
\label{eq:nuOmo}
&& \nu_\Omega=m\left[\frac32-\frac12\zeta^2(y-1)^2\right]~~;
\quad y_0=\zeta(y-1)\le1~~;\quad y\le\frac{1+\zeta}\zeta~~;\qquad
\end{lefteqnarray}
where $\nu_\Omega=3/5$ in the case under discussion.

The reduced mass, $1/m=M_i/M_j$, in terms of the reduced tidal radius,
$1/y_i^\ast=a_i^\ast/a_j$, can be inferred from Eq.\,(\ref{eq:nuOmo}) as:
\begin{lefteqnarray}
\label{eq:myia}
&& \frac1m=\frac53\left[\frac32-\frac12\zeta^2\left(\frac1{y_i^\ast}\right)^
{-2}\left(1-\frac1{y_i^\ast}\right)^{2}\right];~1\ge\frac1{y_i^\ast}\ge
\frac\zeta{1+\zeta}~;\quad0\le\zeta\le1;\qquad
\end{lefteqnarray}
where $1/m=1/\nu_\Omega=5/3$ as $1/y_i^\ast=\zeta/(1+\zeta)$, $1/m=5/2$ as
$1/y_i^\ast=1$, and $y_i^\ast\ge1$ by definition.

In the special case of a globular cluster $(i={\rm C})$ within the Galaxy
$(j={\rm G})$, Eq.\,(\ref{eq:myia}) reduces to (\ref{eq:eoiyi}).

\subsection{von Hoerner's tidal radius}
\label{a:VHtr}

Let a heterogeneous sphere, centered on ${\sf O}_i$, be subjected to the
gravitational force from a mass point placed on ${\sf O}_j$ outside the
boundary.   Let $a_i$ be the radius of the sphere and $R_0$ the distance
$\overline{{\sf O}_j{\sf O}_i}$, where $R_0>a_i$ in the case under
consideration.   Let ${\sf P}$ be the intersection point between the boundary
and the segment, $\overline{{\sf O}_j{\sf O}_i}$.

The gravitational force from the mass point on a unit mass placed on
${\sf O}_i$ and ${\sf P}$, respectively, is:
\begin{lefteqnarray}
\label{eq:FGjH}
&& F_{{\rm G},j}({\sf O}_i)=-\frac{GM_j}{R_0^2}~~;\qquad
F_{{\rm G},j}({\sf P})=-\frac{GM_j}{(R_0-a_i)^2}~~;
\end{lefteqnarray}
and the gravitational force from the heterogeneous sphere on a unit mass
placed on ${\sf P}$ is:
\begin{lefteqnarray}
\label{eq:FGiH}
&& F_{{\rm G},i}({\sf P})=\frac{GM_i}{a_i^2}~~;
\end{lefteqnarray}
which has an opposite orientation wih respect to $F_{{\rm G},j}$.

Accordingly, a local criterion for the definition of tidal radius reads:
\linebreak
$-F_{{\rm G},j}({\sf O}_i)+F_{{\rm G},j}({\sf P})+F_{{\rm G},i}({\sf P})=0$
which, by use of Eqs.\,(\ref{eq:FGjH})-(\ref{eq:FGiH}), after little algebra
takes the explicit expression:
\begin{lefteqnarray}
\label{eq:rtH}
&& \frac{a_i^2}{R_0^2}\frac{2R_0a_i-a_i^2}{(R_0-a_i)^2}=\frac{M_i}{M_j}~~;
\end{lefteqnarray}
where the limit, $R_0\gg a_i$, yields the classical result (von Hoerner 1958).

The reduced mass, $1/m=M_i/M_j$, in terms of the reduced tidal radius,
$1/y_i^\ast=a_i^\ast/a_j$, can be inferred from Eq.\,(\ref{eq:rtH}) via
(\ref{eq:my}), (\ref{eq:y0}), as:
\begin{lefteqnarray}
\label{eq:myiaH}
&& \frac1m=\frac1{\zeta^3}\left[\frac1{y_i^\ast}\left(1-\frac1{y_i^\ast}
\right)^{-1}\right]^3\left[2-\frac1\zeta\frac1{y_i^\ast}\left(1-\frac1
{y_i^\ast}\right)^{-1}\right] \nonumber \\
&& \phantom{\frac1m=}\times
\left[1-\frac1\zeta\frac1{y_i^\ast}\left(1-\frac1{y_i^\ast}\right)^{-1}\right]
^{-2}~~;\qquad y_i^\ast>1~~;\qquad0\le\zeta\le1~~;\qquad
\end{lefteqnarray}
where $1/m\to-\infty$ as $1/y_i^\ast\to1^-$, $1/m\to0$ as $1/y_i^\ast\to0$,
and $1/m=0$ as
$1/y_i^\ast=2\zeta/(1+2\zeta)$.  Keeping in mind $m\ge0$ by definition, the
domain of the function, expressed by Eq.\,(\ref{eq:myiaH}) for assigned
$\zeta$, reads:
\begin{lefteqnarray}
\label{eq:domf}
&& 0<\frac1{y_i^\ast}\le\frac{2\zeta}{1+2\zeta}~~;\qquad0\le\zeta\le1~~;
\end{lefteqnarray}
while, on the other hand, the condition that the mass point lies outside the
sphere implies $y_0>1$ or:
\begin{lefteqnarray}
\label{eq:conf}
&& \frac1{y_i^\ast}<\frac{\zeta}{1+\zeta}\le\frac{2\zeta}{1+2\zeta}~~;
\end{lefteqnarray}
where $1/m\to+\infty$ as $1/y_i^\ast\to\zeta/(1+\zeta)$, and the domain
under discussion reduces to:
\begin{lefteqnarray}
\label{eq:domi}
&& 0<\frac1{y_i^\ast}<\frac{\zeta}{1+\zeta}~~;\qquad0\le\zeta\le1~~;
\end{lefteqnarray}
conformly to Eqs.\,(\ref{eq:domf}) and (\ref{eq:conf}).

The comparison of Eq.\,(\ref{eq:myiaH}), inferred using a classical local
criterion for the definition of tidal radius (von Hoerner 1958) with its
counterpart inferred using a global criterion, Eq.\,(\ref{eq:cuse}) for a
homogeneous sphere, discloses two main differences, namely (i) the tidal
radius is independent of the density profile in the former case but the
contrary holds in the latter and, (ii) the reduced mass, $1/m$, exhibits a
cubic dependence on the reduced tidal radius, $1/y_i^\ast$, in the former
case and a linear dependence in the latter, provided $y_i^\ast\ll1$.
\end{appendix}

\end{document}